# Quantifying Mental Health on Social Media: A Survey

MUSKAN GARG, Thapar Institute of Engineering & Technology, India

Amid lockdown period more people express their feelings over social media platforms due to closed third-place and academic researchers have witnessed strong associations between the mental healthcare and social media posts. The stress for a brief period may lead to clinical depressions and the long-lasting traits of prevailing depressions can be life threatening with suicidal ideation as the possible outcome. The increasing concern towards the rise in number of suicide cases is because it is one of the leading cause of premature but preventable death. Recent studies have shown that mining social media data has helped in quantifying the suicidal tendency of users at risk. This potential manuscript elucidates the taxonomy of mental healthcare and highlights some recent attempts in examining the potential of quantifying suicidal tendency on social media data. This manuscript presents the classification of heterogeneous features from social media data and handling feature vector representation. Aiming to identify the new research directions and advances in the development of Machine Learning (ML) and Deep Learning (DL) based models, a quantitative synthesis and a qualitative review was carried out with corpus of over 77 potential research articles related to stress, depression and suicide risk from 2013 to 2021.

## 1 BACKGROUND

According to World Health Organization, every year there are more than 0.8 million people who die out of suicide. According to the Centers for Disease Control and Prevention (CDC) WISQARS, *Leading Causes of Death Reports*, a report of 2019, *suicide* was the tenth leading cause of death in United States. According to the official data of USA, one person commits suicide in every 11.1 minutes[1]. According to the latest available data, the statistics of Canada estimates 4,157 suicides in 2017, making it the ninth leading cause of death. The clinical psychologists and academic researchers have come across increasing number of mental health problems and its exposure to the social media platforms during the COVID-19 pandemic lockdown. The pandemic has long-term impacts on the mental health and wellness of masses due to economic insecurity and isolation. The suicide cases have adverse physical, economical, and emotional impact on the well-being of families and communities. Early suicide risk prediction may help to control the suicide rate by taking preventive measures and good governance. The digital data is identified as the most significant data for early prediction of suicidal risk and thus, help in promoting well-being of the global mental health.

[1]https://suicidology.org/wp-content/uploads/2021/01/2019datapgsv2b.pdf

The social media users generate first-hand information in the form of *post* on social media about their thoughts, beliefs and emotions amid pandemic lockdown situation. The post on social media may contain heterogeneous type of unstructured and ill-formed data which is appropriate for manual intervention but are difficult to decode automatically by a system. Thus, the automatically accessing and processing social media multi-modal data for quantifying the suicidal tendency in *post* is one of the leading research domain in mental healthcare. Recent studies on predicting suicidal tendency on social media data by using machine learning [4, 5, 9, 90, 112] models are more successful as compared to the medical records [26] and paved the way to explore deep learning [10, 58, 75, 93, 107, 134] and computational intelligence techniques[101] for quantifying suicidal tendency.

Other than stress, clinical depression, suicide risk, studies over six other social mental health problems are observed in literature. However, the other six studies are not directly associated with the suicidal tendency and hence, is out of the scope of this manuscript. Moreover, among nine mental health problems, stress, clinical depression and suicidal tendency detection on social media are the most widely studied areas.

### 1.1 Motivation

The vanilla clinical psychology study is the theoretical approach which needs manual labour for identifying the suicidal tendency. This labour intensive engineering is subjective and follows the time consuming face-to-face interaction. The major challenge of traditional psychological sessions is that 80% of people who are at risk are not comfortable in disclosing the level of stress and anxiety that they may have [62]. In this context, it becomes important to identify people at risk to control the suicide rate which is believed to have been preventable [110].

With this background, the studies over quantifying the suicidal tendency may help in good governance of the country by identifying the locations where the suicidal ideation is trending on social media. In the medical domain, it is important to predict the available and required resources in near future for medical assistance to mental illness detection, prevention and control. This automatic mental health predictions helps in reducing the labour and cost of identifying people at risk. The social media platforms like Facebook has used suicide risk detection model for identifying potential users at risk and offer them help [118].

Researchers have found that social media have strong association with expression of feelings by users [13, 38, 57] and about eight out of ten people tend to disclose their suicidal tendencies [35] on social media. The mental health prediction from social media [20] has helped in suicidal risk assessments [104]. An extensive literature survey on predicting suicidal tendency may enable the academic researchers to enrich this domain by identifying new research directions from this manuscript.

### 1.2 Mental Healthcare: A taxonomy

The social mental health has evolved as a unique body of knowledge and the taxonomy of mental healthcare is shown in Fig. 1, Mental healthcare is recently expanded to computational linguistics and human-computer interaction for integrated research to automate the predictions. Conventionally, there are two ways to identify people at risk, the digital data and the traditional offline interactions. As discussed earlier, the use of traditional method is reduced as people refrain themselves from visiting clinical psychologists.

The *biomedical* domain deals with neuroscience based investigations, clinical reports and diagnosis using EEG signals. The *social aspect* of mental health studies is closely associated with the human-behaviour within the society. The *psychological aspect* is more inclined towards theorizing the thoughts on mental health. The *ethical aspect* is concerned exclusively with the security of the data, to what extent and in what manner it can be used [12]. The *prevention and control* measures for mental health issues can be taken after *identifying or predicting* the people at risk.

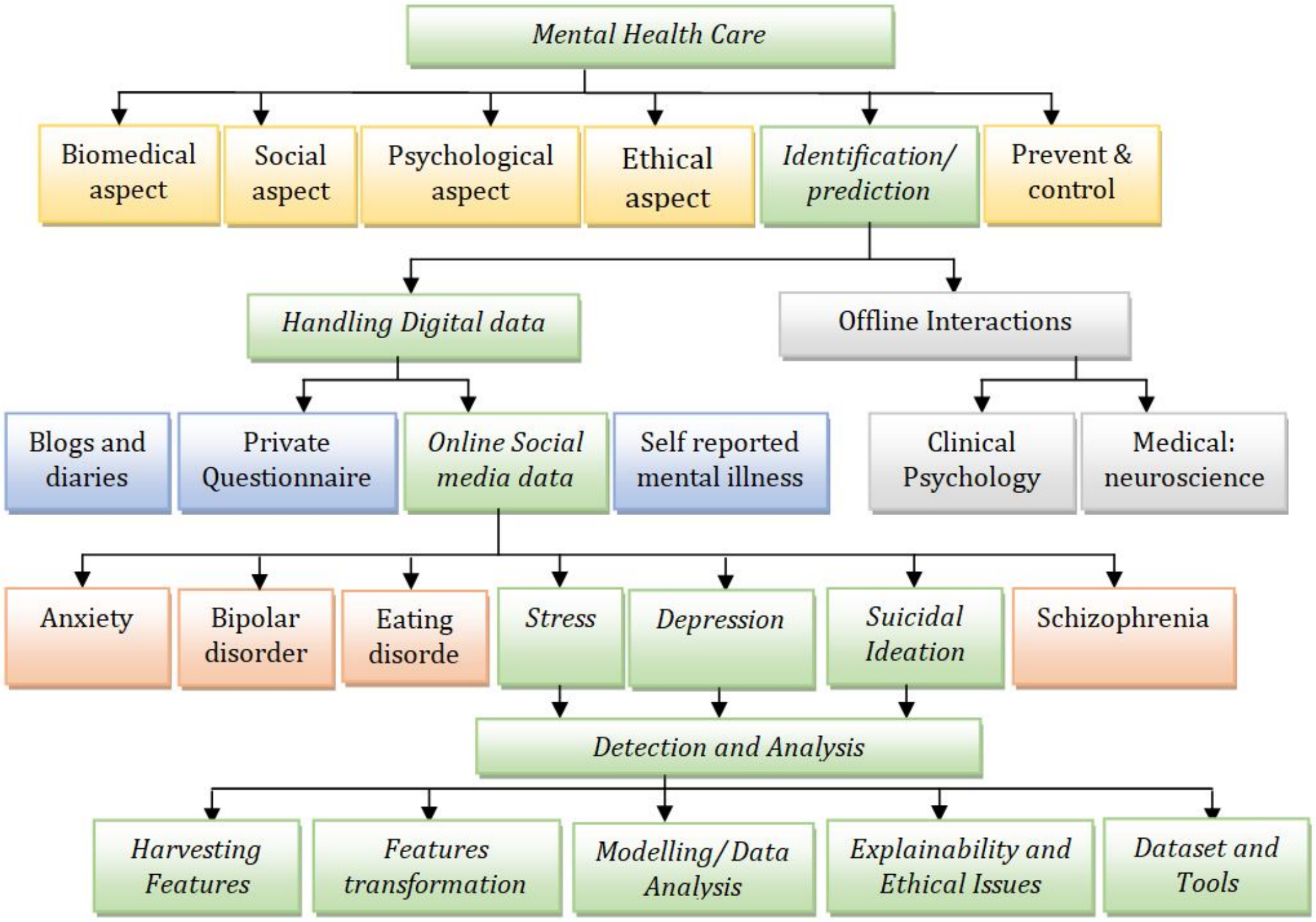

Fig. 1. Taxonomy on Mental Healthcare

The investigations on digital mental health has shown many significant studies over *blogs and diaries* of the user, information filled in *private questionnaires* or Google forms, *self-reported mental illness* voluntarily, and *online social media data*. The online social media contains heterogeneous data, like, linguistic information, user-information, social information, and multimedia data. The scope of this manuscript is to deal with textual social media platforms (Twitter, Reddit, Sina Weibo) which occasionally contains images for stress, depression and suicide risk on social media. There are studies over multimedia (images, audio and visual) data based social media platforms (Instagram, Youtube[2]) in existing literature which is beyond the scope of this manuscript due to different nature and semantics of the metadata.

### 1.3 Corpus Overview

After extensive literature survey, among 77 research articles, 7 articles related to *stress*; 28 articles related to *depression*; 30 articles are related to *suicide risk*; 12 articles related to two or more *mental health* disorders, are examined in detail. The year-wise distribution of number of research articles is shown in Fig. 2 which are published in top venues like CLPsych (8), ACL (7), and AAAI (4) as shown in Fig. 3.

[2]https://dcapswoz.ict.usc.edu/

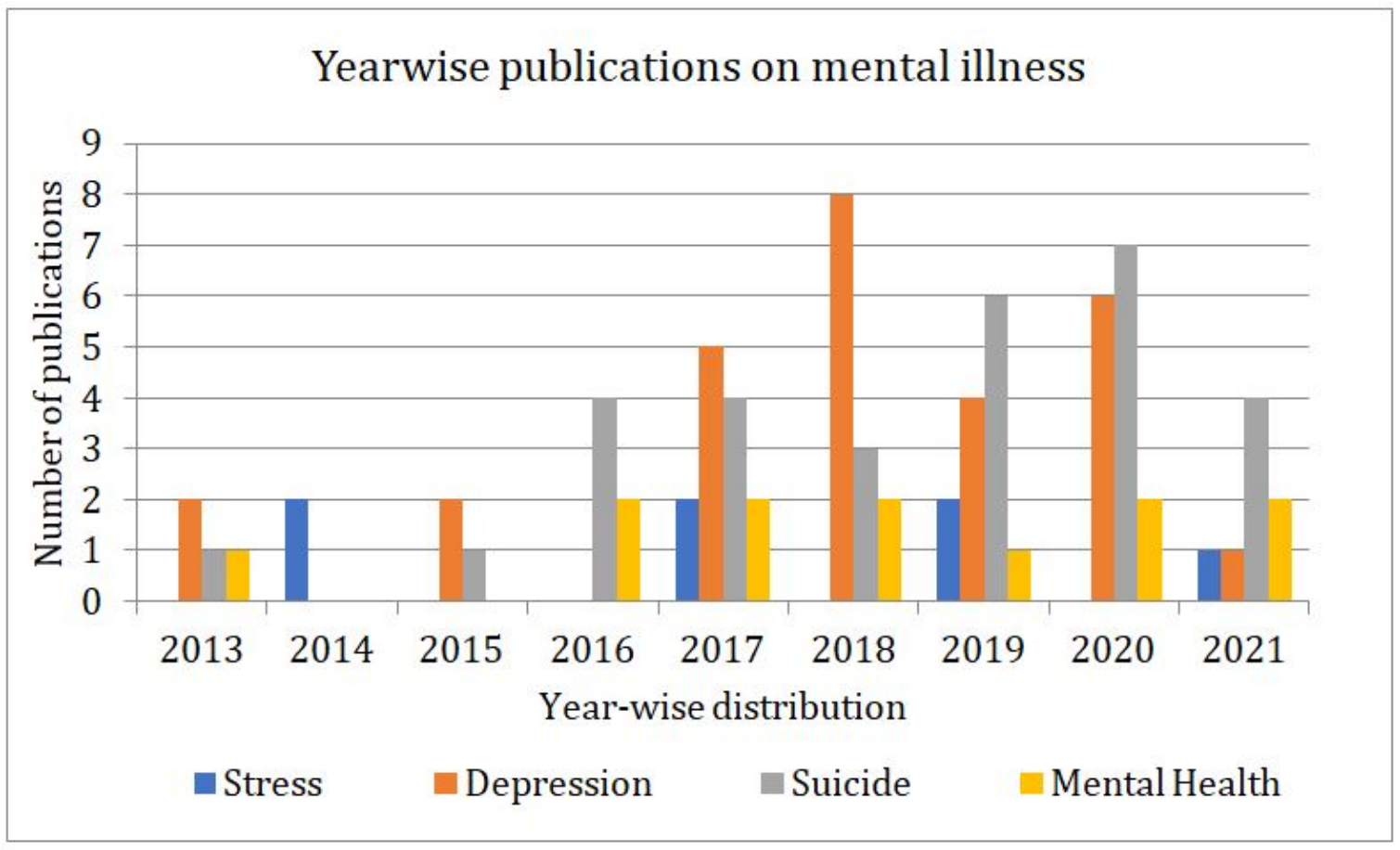

Fig. 2. Year-wise distribution of number of articles on mental disorders

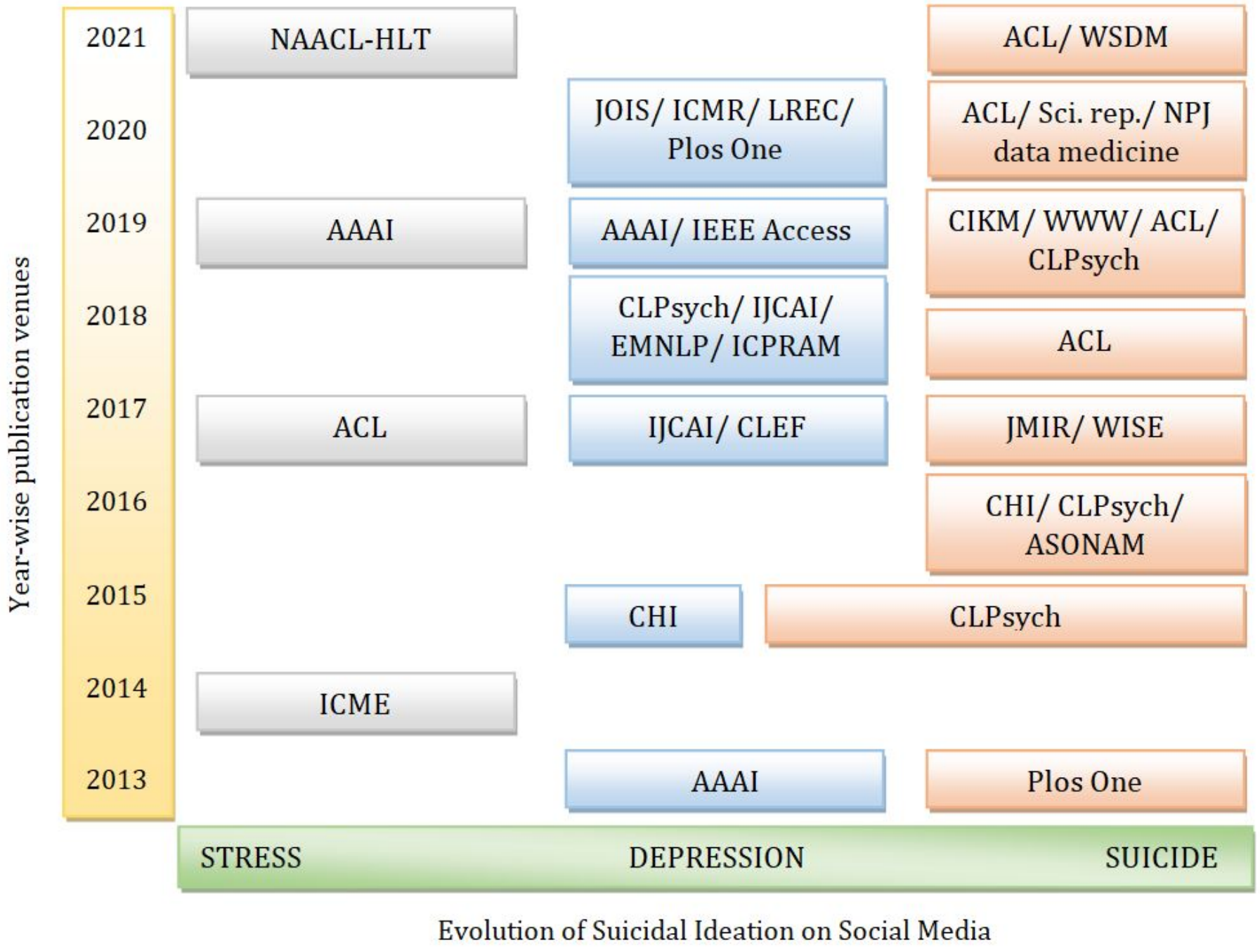

Fig. 3. Year-wise distribution of publication venues

The corpus of 77 articles does not include the review or survey articles [12, 18, 38, 47, 125]. The research articles on stress and suicide risk detection are fewer than the article on identifying clinical depression.

Initially, the multidisciplinary journal, *Plos One* have shown interest for publication of the study associated with the suicide on social media [59] and it is still evolving in recent years [130]. The area of interest by research community has evolved from *social venues*, [19, 65, 83, 88] to *Human-Computer Interaction* venues, [22, 114] and *Computational Linguistic* domain of computer science [94, 128]. Recently, a highly reputed publication house, *Nature*, has validated this as an interesting body of knowledge [75]. Existing studies have addressed the concerns on dataset and its ethical constraints [24, 38, 39, 115]; multi-modal feature extraction [36, 50, 60, 102]; classification techniques [4, 5, 9, 58, 102, 111, 112, 133], graph learning approach [9]; use of the Internet of Medical Things for real-time applications[39]; noisy label problem in dataset annotations[40]; and improvement over the attention mechanisms [17, 58, 107, 135].

### 1.4 Scope of the study

The research domain of Mental Illness Detection and Analysis on Social media (MIDAS) has evolved for less than a decade [21]. There are a few survey and review articles on MIDAS. The existing studies on public opinion [29] shows that the participants could visualize possible benefits of this study but not at the cost of privacy concerns. An in-depth study about the dataset and its ethical issues were explored in a systematic review in 2017 for statistical analysis of mental health dataset[38]. A Critical review of 75 research articles on the the mental health issues from 2013 to 2018 was given to study the design and research methods [13]. A short survey was found to address some concerns of the connectedness of social media data and mental health prediction [18]. Some recent studies were made for feature extraction and online behaviour patterns used for mental health prediction using recent deep learning techniques [47].

Some well-established studies over the generalized analysis of mental health detection [7, 12, 113, 125] have set the foundation for this potential field of study. To investigate the specific mental health problems, a survey on identifying clinical depression [32, 70] were observed in the literature due to extensive research on depression detection. However, this survey article is focused on the studies which are directly associated with suicide risk and the extent of suicidal tendency and can be used for real-time applications to control the suicide risk. After extensive literature survey of 77 research articles, this manuscript addresses the following concerns:

- Classification of heterogeneous social Media Features
- State-of-the-art on evolution of multi-modal data fusion and classification techniques
- Enumerating the tools, resources, and available dataset
- Highlights the open challenges and new research directions.

This organization of this manuscript is structured as follows. Section 2 presents the classification of different features extracted from social media data for suicide risk detection. The feature extraction, feature selection and post feature transformation modules are elaborated with existing literature. Section 3 summarizes the evolution of automated learning based techniques for quantifying stress, depression and suicide risk on social media. The available dataset are enumerated and other tools and resources were explored. Section 4 highlights the open challenges and new research directions and gives the glimpse of future scope of study. Finally, Section 5 concludes the manuscript.

## 2 FEATURES FROM SOCIAL MEDIA DATA

With this background, some interesting research questions arise related to the *data curation* which is one of the most challenging task to handle social media data due to its unstructured/ semi-structured, user-generated and ill-formed nature. The social media data contains heterogeneous information. The text classification [66] is one of the most promising research areas which is often used for classifying mental health illness on social media. Few existing studies exclusively deals with feature extraction from social media data for mental health illness [101, 109] and have addressed the concerns to explore dominant features [43]. Thus, the vector space representation of social media data is discussed in this Section.

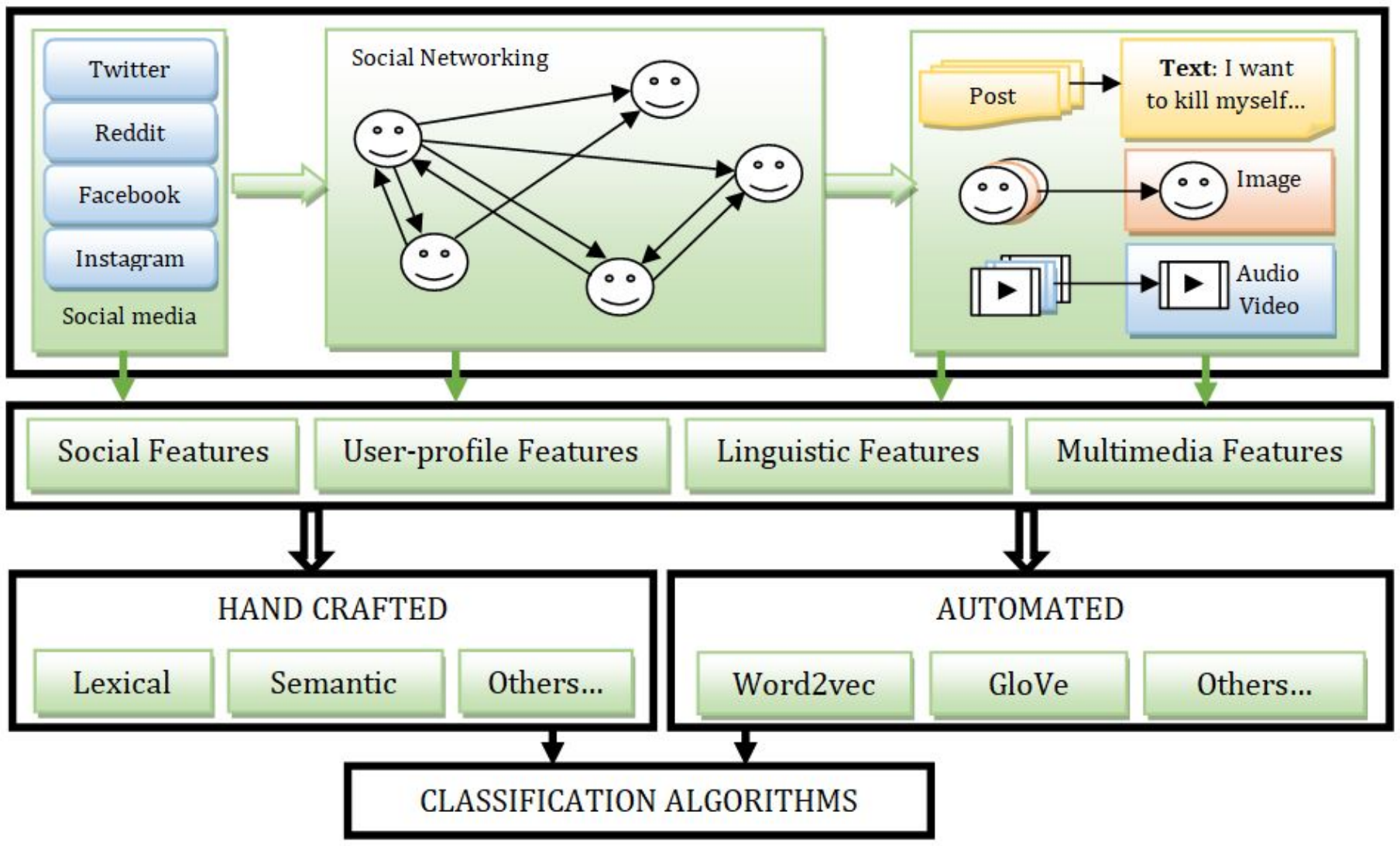

Fig. 4. Architecture of feature harvesting from social media data for classification algorithms

The social media platforms are usually characterized by one-way connections (Twitter, Reddit, Instagram) and two-way connections (Facebook). The most widely used social media platforms are Twitter [36, 47, 51, 53, 96, 97] and Reddit [40, 60, 76, 107], followed by Instagram [85, 126], and Facebook [26, 37, 75]. The social media platforms contains different modalities of data, viz., *text*, *image*, *multimedia* which is transformed to the vector space for mining useful insights. The length of *text* in a tweet of the Twitter social media platform is not more than 280 characters [34].

The number of unique words in the dataset remains limited due to limited amount of information and limited words in the active vocabulary of users. This limited information contains patterns among them which are identified for classifying the data. These patterns are identified by feature extraction and transformation for mining information to classify the data.

## 2.1 Classification of Features

The statistical studies over feature extraction has given many useful insights about behavioural analysis over social media data [43, 56, 109, 114, 117]. The architecture of harvesting features from social media data for classification problem is shown in Fig. 4.

The social data is broadly classified into two modalities *textual information* which are regularly posted and *images* which are occasionally posted on social media platforms. The textual data is given as an input for feature extraction using either the conventional approach of *Vectorizer* or the *embedding* techniques. The embedding are further classified into *static embedding* and *dynamic embedding*.

The social media features are extracted and classified into *handcrafted features*, the statistical information, and *automated features*, the automation on attributes of data. Both handcrafted features and automated features are given as an input to the classification algorithms to obtain suicidal tendencies. The social media data is broadly classified into four different categories to discuss the features for suicidal tendency identification, namely, user-profile features, linguistic features, social features, and multimedia features as shown in Fig. 6.

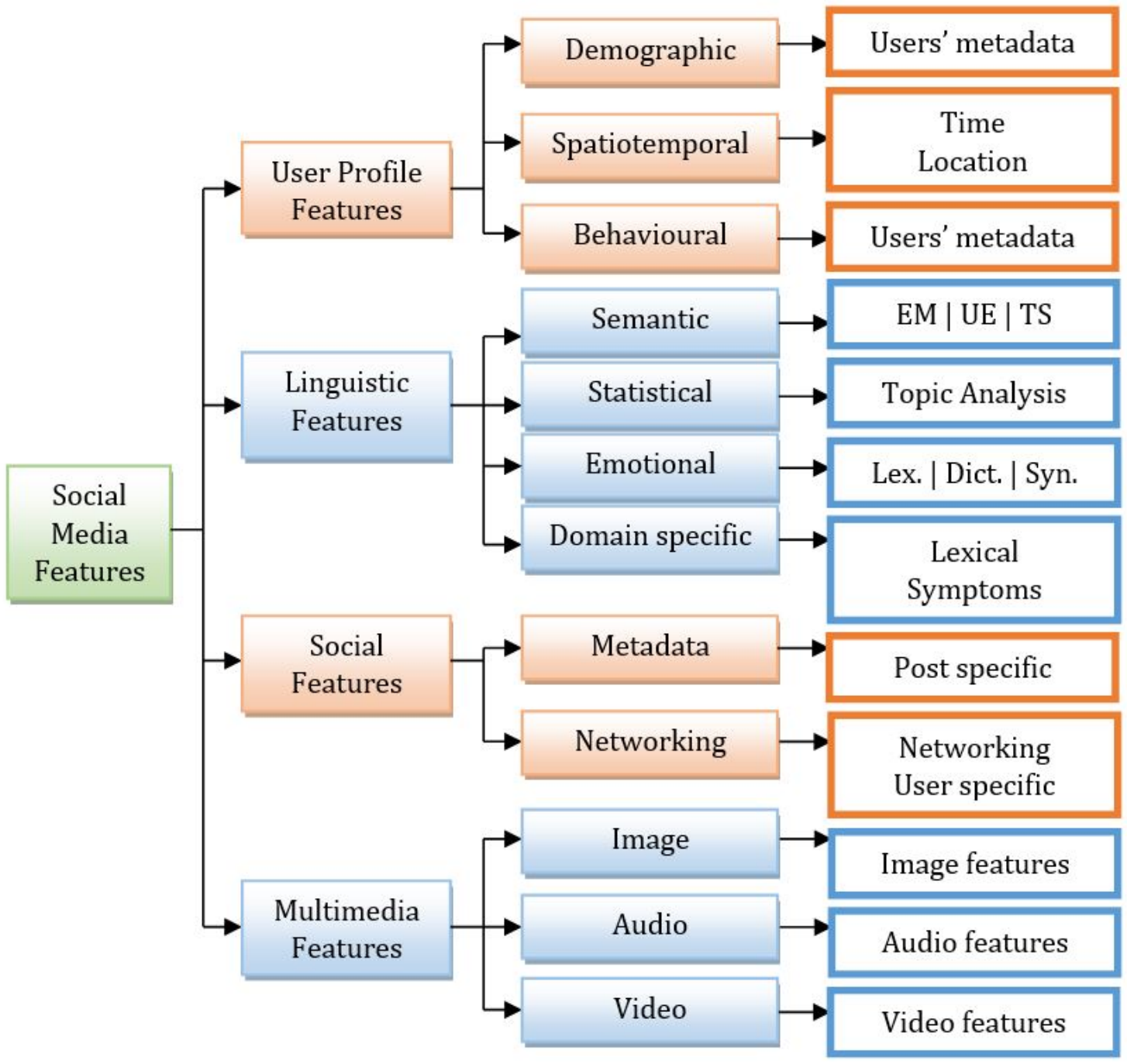

Fig. 5. Classification of social media features for quantifying suicidal tendencies

### 2.1.1 *Handling Ambiguity of Features.*

Although there is no ideal classification of features, the social media features are classified into four different categories and few exceptions belong to two or more different categories.

- In addition to the textual data from posts, the metadata of social media platforms is examined to extract more features. The metadata is classified into the *user metadata*: data about the users' profile and is thus, kept under *user profile features*; and *post metadata*: data about the post and is categorized under *Social features*.
- The ruminative response style can be expressed as repetitive thoughts and behavior [73]. People with depression tend to express their feelings or negative experiences repeatedly so that sentences in relevant topics may also repeatedly appear on their posts. Though the ruminative response style is the part of both user behaviour and linguistic styles, it is more closely associated with user-profile features and thus, studied under *User Profile Feature*.
- An interesting study introduced bBridge [28], a big data based feature extraction approach from social media data which both contains user-profile features and social networking features.
- The community specific information of the user comprises of the information about followers, and favourites. Since, these features are more associated with the social networking of a user, they are discussed in the *Social Features*.

| Category | Sub-Category | Feature | Type | Research Articles |
|---|---|---|---|---|
| Demographic | Users' Meta-data | Age | User | [28, 59, 60, 102] |
| | | Gender | User | [28, 59, 60, 60, 102] |
| | | Education | User | [28, 102] |
| | | Occupation | User | [28, 102] |
| | Users' network | Follower | User | [67] |
| | | Ego-network | User | [6] |
| Spatio-temporal | Temporal | Timeline | User | [60, 93, 102] |
| | Spatial | Location | User | - |
| Behavioural | Posting Behavior | Gen. Behaviour | User | [102] |
| | | Ruminative | User | [107] |
| | | Posting Time | User | [37, 103] |
| | | User Emo. | User | [37, 38, 50, 51, 53] |

Table 1. User profile feature extraction for mental health status.

*2.1.2 User Profile Features.* Existing studies states that people having a college degree or a regular job are less likely to be depressed [77]. The users sharing the similar demographic, linguistic and cultural traits are found to be more at risk than others [82]. The user-profile features are shown in Table 1.

*Demographic*: The *users' metadata* of user contains information about the age [82], gender, sometimes educational details and occupation.

*Spatio-temporal*: The *PHASE* of a user is mapped by tracking its historical tweets and convolution approach across these tweets[93]. Existing studies have identified the patterns of irregularities among posting behaviour. The location dimension may have strong associations of economical indexes like *Ease of doing business*[3] and *World Happiness Report*[4] with mental health status of residents.

*Behavioural features*: The depressed tweets are more likely to be expressed late night [102] than during day time. The *insomnia index* or the sleep cycle [21] and repetitive thoughts and behavior [73] of the user affects the mental health. People with depression tend to express their feelings or negative experiences repeatedly.

*2.1.3 Linguistic Features.* The authors have identified the importance of words that users pick to express their feelings of depression [77] in their existing studies. People with depression exhibit differences with respect to linguistic styles such as the distribution of nouns, verbs and adverbs and the complexity of sentences, which are conceptualized unconsciously [33]. There are few studies exclusively on linguistic features [74] and identified the increased use of first person language, the current scenario and anger based terms. The linguistic features are further classified as shown in Table 2

*Emotional Features*: The emotional models are used as features in existing studies [60]. Recently, the information about emotion models is summarized due to evolving era of 'Emotional Artificial Intelligence' for affective computing [131]. The Valence arousal dominance (VAD) Emotion model [60, 98, 102, 135] and Plutchik [81, 93] were used for social mental health prediction. The Emonet was [1] issued to fine-tune the BERT model like Plutchik Transformer [93] evolved from the Plutchik's wheel of emotions [81].

[3]https://en.wikipedia.org/wiki/Ease_of_doing_business_index
[4]https://en.wikipedia.org/wiki/World_Happiness_Report

| Category | Sub-Category | Feature | Type | Research Articles |
|---|---|---|---|---|
| Emotional | Emotional Model | Plutchik | Post | [93] |
| | | VAD | Post | [48, 60, 102] |
| | | Affect & Intensity | Post | [60] |
| | | big 5 personality [99] | Post | [60] |
| | | Anxiety, anger, dep. | Post | [60] |
| | Textual Sentiments | Emoji | Post | [51, 53, 102, 103] |
| | | Emoticons | Post | [51, 53, 103] |
| | | SentiWordNet | Post | [107] |
| | | SentiNet [8] | Post | [4] |
| Semantic | Topic Analysis | LDA | Post | [37, 60, 61, 102, 111] |
| | | Brown Clustering | Post | [68] |
| Statistical | Lexical | TFIDF | Post | [92, 93, 96, 112] |
| | | Text | Post | [40] [36, 76, 96, 103, 116] |
| | | Morphological | Post | [109] |
| | | Stylometric | Post | [109] |
| | | n-gram | Post | [4, 22, 37, 60, 96, 101, 109, 111] |
| | | Punctuation | Post | [51, 53] |
| | Dictionary | LIWC[5] | Post | [21, 22, 37, 96, 101, 102, 111] |
| | | ANEW[6] | Post | [71] |
| | Syntactical | POS Tagging | Post | [4, 22, 45, 96, 103, 107, 109] |
| Domain Specific | Lexicon | Antidepressant | Wiki | [4, 102, 107] |
| | | TensiStrength | Dict. | [37] |
| | | Dictionaries | Wiki | [4] |
| | Dep. symptoms | DSM[124] | Dict. | [102, 107] |

Table 2. Linguistic feature extraction for mental health status

*Semantic Features*: The topic modelling method, *LDA*, [88] [68] is used for clustering the posts related to similar topics. The depressed and non-depressed users discuss different topics [87] which may help to determine potential depressed users.

*Statistical Features*: The statistical features are categorized into lexical features, dictionary based features, and syntactical features. The *lexical features* use tokenized form of the text to calculate statistical measures like TFIDF, n-grams, morphology and alike features. The *Dictionary features* are associated with the use of existing dictionaries for assigning values like LIWC and ANEW. The *syntactical features* are used to check the context of the token with respect to the neighbouring words, for instance, Part-Of-Speech tagging.

*Domain Specific*: The domain specific features are the lexicon of mental health specific words derived from Wikipedia, domain specific dictionaries, and depression symptoms (DSM). The contribution of domain specific features significantly improve the results due to contextual association of words.

*2.1.4 Social Features.* People who are depressed and share their feelings on social media platforms like Twitter are very conscious about their social circle on social media and have very limited number of friends [78]. The depressed tweet gains more attention from friends and so, the important features includes Retweets, comments, and favourites [51]. The social features are further classified into *social metadata* and *social networking* as shown in Table 3.

*Social Metadata*: The information about the post of the user consists of the length of a post, number of hashtags in a post, number of URLs used in a post and other minute details which is termed as the metadata.

| Category | Sub-Category | Feature | Type | Research Articles |
|---|---|---|---|---|
| Social metadata | Post Specific | Length | Post | [14, 22, 93, 135] |
| | | #(Hashtags) | Post | [37] |
| | | #(URL) | Post | [37] |
| | | Metadata | Reddit | [60] |
| Social Network | Networking | Interactions | User | [22, 50, 50, 51, 53, 59, 102, 103, 135] |
| | | At-Mentions | User | [67] |
| | | Replied to | User | [67] |
| | User Specific | #(Favourites) | User | [51, 103] |
| | | #(Likes) | User | [53] |
| | | #(Posts) | User | [102, 103] |
| | | #(Comments) | User | [51, 52] |
| | | #(ReTweet) | User | [51, 92, 135] |

Table 3. Social feature extraction for mental health status.

| Category | Feature | Type | Research Articles |
|---|---|---|---|
| Image | Colour Combinations | Image | [36, 50, 51, 102] |
| | Colour Ratio | Image | [36, 50, 51, 102, 103] |
| | Brightness | Image | [36, 50–52, 102, 103] |
| | Saturation | Image | [36, 50, 51, 53, 102, 103] |
| | Convolution | Image | [36, 122, 132] |

Table 4. Multimedia feature extraction for mental health status detection

*Social Network*: Few interesting studies are carried out to obtain patterns from the interaction and relationships of users [95]. These networking features are used for graphical neural network approach.

*2.1.5 Multimedia Features.* The *images* which are considered for feature extraction or transformation are either the display picture in Twitter (also referred as Avtars in Reddit) or the image posted by a user. The colour combinations, colour ratio, brightness, saturation, and convolution are used for mining social media data as observed in Table 4.

## 2.2 Feature Vector Representation

The feature vectorization is the process of feature representation as an input to the classification algorithms. Feature vector representation is further classified into textual feature vectorization and image feature vectorization as shown in Fig. 6. The *textual feature vectorization* is comprised of feature extraction and feature embedding and existing literature are classified for feature vector representation in Table 5.

*2.2.1 Textual Feature Extraction.* Earlier, the textual features (TFEx) are converted into vectors by using conventional approach of TFIDF vectorizer, Count vectorizer, and Hashing vectorizer [40]. For dimensionality reduction, the selective features are processed further by using PCA, NMF and other filter based linear feature selection algorithms. In existing literature, authors have used one-hot encoding to encode the set of Tweets [76]. The uni-modal dictionaries are generated from textual and image data separately which are further used for joint sparse representation [102]. These traditional feature extraction techniques are used for converting the social media data into vector representation for classification models.

*2.2.2 Feature Embedding.* With advancements in the word to vector conversion using neural network approach, the word2vec [64], the GloVe [10, 107], and the Fasttext were used for feature vector representation. To handle the longer text like phrase, sentence or paragraph, the authors used BERT [25], Sentence-BERT [86], and Google Universal Sentence Encoder (GUSE) [11] for feature vector representation [40]. Few studies have used embedding over dense layers, BERT, GUSE, and GRU [36, 107] for sequence to sequence learning and the use of attention enhance the importance of feature across this representation.

An image represents many characteristics of the psychological thoughts and health. The permutations and combinations of different image features extraction helps in determining the mental health. To use image as features, the authors have followed end to end feature transformation technique by using a 16-layer pre-trained VGGNet [36, 105, 132].

*2.2.3 Dimensionality Reduction.* One of the most promising step of social media data mining is dimensionality reduction. The dimensions of textual representation in conventional feature extraction techniques are reduced by linear and non-linear methods which are comprised of Principal Component Analysis (PCA), Deep Neural Autoencoders (DNAE) [121], and Uniform Manifold Approximation and Projection (UMAP) [63] for dimensionality reduction in MIDAS.

The Post Feature Tranformation (PFT) approach is used for transformer embedding based end to end conversion of data into feature vectors. Authors have used an attention mechanism like Hierarchical Attention Mechanism (HAM) [128] [44] to give importance to important posts [135] [107] for identifying suicidal tendencies.

In few studies, the authors have used the multimodal dictionary learning and joint sparse representation [102] for feature transformation. Another path breaking work in literature on multi-attributed feature extraction was introduced with 3-level framework [51]. A three level framework was proposed by using three-level features extraction which consists of low level feature (linguistic features), middle level features (visual features) and high-level features (social features) to give as an input to the Deep Sparse Neural Network (DSNN). Authors argued that all the three types of features are not always available and thus, the Cross media Auto Encoder (CAE) was proposed for joint representation using Denoising Auto-Encoder (DAE).

## 2.3 Summary of Feature Extraction and Transformation

After extensive literature survey, few interesting studies have potential to define and explore new features for mental health detection from social media data as shown in Table 6. Most of the recent approaches use embedding techniques and work on post feature transformation to hypothesise the better feature representation. Moreover, all the existing studies are using the textual information of post and other features optionally.

| **Features** | **Category** | **Subcategory** | **Type** | **Papers** |
|---|---|---|---|---|
| Traditional FE | Statistical | Vectorizer | Post | [2, 40, 101, 109] |
| | | Entropy | Post | [21] |
| | | Statistics | Post | [21] |
| | Dictionaries | Dict. learning | Post | [102, 103] |
| Embedding | Static | Word2vec [64] | Post | [10, 76, 83, 107, 109, 112, 132] |
| | | GLoVe | Post | [10, 107] |
| | | Fasttext | Post | [10, 107] |
| | CE: Transformer | BERT[7] [25] | Post | [50, 60, 93] |
| | | Sentence BERT[86] | Post | [40, 60] |
| | | GUSE[11] | Post | [40] |
| | Encoding Seq. | RNN[27] | Post | [107, 107] |
| | | GRU[15] | Post | [36, 135] |
| | | LSTM[42] | Post | [93, 103] |
| Image FE | Image | VGGNet[105] | Image | [36] [132] |
| | | ImageNet[91] | Image | [132] |
| | | CNN | Image | [50] |
| Dim. Red. | Linear | Filter | Vectors | [21, 40, 121] |
| | Non-linear | NMF | Vectors | [60] |
| | | t-SNE | Vectors | [111, 112] |
| | Post Feature T. | HAN[8] [128] | Vectors | [17, 44, 107, 120, 135] |
| | | Joint Sparse Repr. | Vectors | [51, 53, 93, 102, 103] |
| | | Optimization | Vectors | [76, 101] |

Table 5. Feature vector representation for social mental status detection

| **Paper** | **Year** | **F1** | **F2** | **F3** | **F4** | **TFE** | **Emb.** | **DR** | **Output** |
|---|---|---|---|---|---|---|---|---|---|
| Choudhury *et. al.,* [21] | 2013 | ✓ | ✓ | ✓ | | ✓ | ✓ | | Depression |
| Lin *et. al.,* [51] | 2014 | | ✓ | ✓ | ✓ | | | ✓ | Stress |
| Lin *et. al.,* [53] | 2017 | | ✓ | ✓ | ✓ | | ✓ | ✓ | Stress |
| Shen *et. al.,* [102] | 2017 | ✓ | ✓ | ✓ | ✓ | ✓ | | | Depression |
| Song *et. al.,* [107] | 2018 | | ✓ | | | ✓ | ✓ | ✓ | Depression |
| Sawhney *et. al.,* [96] | 2018 | | ✓ | | | | ✓ | | Suicidal Id. |
| Orabi *et. al.,* [76] | 2018 | | ✓ | | | | ✓ | ✓ | Depression |
| Tadesse *et. al.,* [111] | 2019 | | ✓ | | | ✓ | | ✓ | Depression |
| Matero *et. al.,* [60] | 2019 | ✓ | ✓ | ✓ | | | ✓ | ✓ | Suicidal Id. |
| Gui *et. al.,* [36] | 2019 | | ✓ | | ✓ | | ✓ | ✓ | Depression |
| Guntuku *et. al.,* [37] | 2019 | ✓ | ✓ | ✓ | | ✓ | ✓ | | Stress |
| Xu *et. al.,* [127] | 2020 | ✓ | ✓ | ✓ | ✓ | ✓ | | | Mental Health |
| Lin *et. al.,* [50] | 2020 | | ✓ | ✓ | ✓ | | ✓ | | Depression |
| Sawhney *et. al.,* [93] | 2021 | ✓ | ✓ | ✓ | ✓ | | ✓ | ✓ | Suicidal Id. |
| Haque *et. al.,* [40] | 2021 | | ✓ | | | ✓ | ✓ | ✓ | Suicide & Dep. |
| Zogan *et. al.,* [135] | 2021 | ✓ | ✓ | | | | ✓ | ✓ | Depression |
| Turcan *et. al.,* [116] | 2021 | | ✓ | | | | ✓ | ✓ | Stress |
| Zogan *et. al.,* [133] | 2021 | ✓ | ✓ | ✓ | ✓ | | ✓ | ✓ | Depression |

Table 6. Feature Extraction and Transformation for Mental Health Detection. F1: User profile Feature, F2: Linguistic Feature, F3: Social Feature, F4: Multimedia Feature, FEx: Feature Extraction, DR: Dimensionality Reduction, FEm: Feature Embedding, PFT: Post Feature Transformation

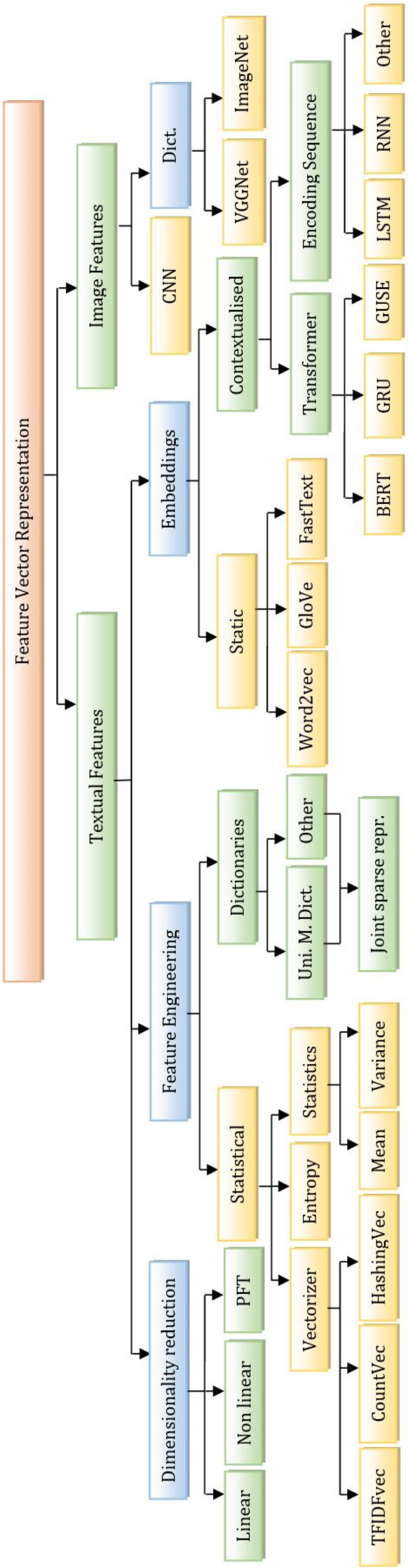

Fig. 6

## 3 CLASSIFICATION

The classification problem of identifying suicidal tendency on social media use many shallow learning and deep learning algorithms. One of the most challenging module is to handle the unstructured and semi-structured data from social media data, filling missing values and jointly represent the multi-modal information. Although, the data resource for this task is freely available in public domain, most of the dataset are not available due to sensitivity of the data.

### 3.1 Available dataset

Most widely used dataset are CLPsych shared task [19], Reddit Self-reported Depression Diagnosis [129], and Language of Mental Health [33], early risk prediction on the internet (eRisk) [54]. As discussed earlier, only a few dataset are available in public domain for the pilot study and many of them are either reproducible or available on request. Every year more than 12 dataset are introduced for prediction of mental health disorder prediction from social media data due to ethical issues, we shall limit our studies to only the most popular dataset, reproducible dataset, and the dataset which are available by request or via signed agreement to ensure the reproducibility of results for stress, depression and suicide detection from social media data. The list of the dataset which are available for this task are enumerated in Table 7.

The most widely used dataset are MDDL, CLPsych, eRISK and UMD- Reddit dataset. However, the recently introduced dataset are widely used for experiments and evaluation of existing and newly proposed techniques. The dataset which are available in the public domain are discussed in this Section.

*3.1.1 Multimodal Dictionary Learning (MDDL):.* MDDL[10] is a depression detection dataset which comprises of three modules D1, D2, and D3. The *Depression Dataset D1* is constructed using tweets from 2009 and 2016 where users were labeled as depressed if their anchor tweets satisfied the strict pattern "(I'm/ I was/ I am/ I've been) diagnosed depression". The *Non-Depression Dataset D2* is constructed in December 2016, where users were labeled as non-depressed if they had never posted any tweet containing the character string "depress". Although D1 and D2 are well-labeled, the depressed users on D1 are too few, thus, a larger unlabelled *Depression-candidate Dataset D3* is constructed for depression behaviors discovery which contains much more noise.

*3.1.2 SDCNL dataset.* : The SDCNL[11] dataset was collected using Reddit API and scraped from two subreddits, r/SuicideWatch and r/Depression which contains 1,895 total posts. Two fields were utilized from the scraped data: the original text of the post as our inputs, and the subreddit it belongs to as labels. Posts from r/SuicideWatch are labeled as suicidal, and posts from r/Depression are labeled as depressed.

*3.1.3 Reddit Self-reported Depression Diagnosis(RSDD).* : The RSDD dataset[12] contains the Reddit posts of approximately 9,000 users who have claimed to have been diagnosed with depression ("diagnosed users") and approximately 107,000 matched control users. The introduction to Reddit dataset [129] has given a significant contribution which was used by many existing studies.

*3.1.4 Self-Reported Mental Health Diagnoses (SMHD) dataset.* : The SMHD dataset [13], just like RSDD dataset, can be obtained via signed agreement as per the privacy policy of data. The dataset consists of Reddit posts of the users diagnosed with one or several of nine mental health conditions ("diagnosed users"), and matched control users. This dataset is also used by few studies in literature and is related to multiple mental health conditions instead of just the depression dataset.

[10]https://github.com/sunlightsgy/MDDL
[11]https://github.com/ayaanzhaque/SDCNL/tree/main/data
[12]http://ir.cs.georgetown.edu/resources/rsdd.html
[13]http://ir.cs.georgetown.edu/resources/smhd.html

| Dataset | Origin | Details | Nature | Papers | Avail. |
|---|---|---|---|---|---|
| MDDL | Shen *et. al.*,[102] | 300 mn users, 10 bn Tweets D1: Depression D2: Non-dep., D3: Dep. candidate | Depression | [36, 50, 102, 133] | ✓ |
| SDCNL | Haque *et. al.*, [40] | Reddit dataset of 1895 posts of depression and suicide | Suicidal and Depression | [40] | ✓ |
| RSDD | Yates *et. al.*, [129] | Reddit dataset of 9210 users in depression and 1,07,274 users in control group | Depression | [17, 107, 129, 133] | S |
| SMHD | Cohen *et. al.*, [16] | Reddit dataset for multi-task mental health illness detection problem domain | Mental Health | [16, 30] | S |
| SRAR | Gaur *et. al.*, [31] | Composed of 500 Redditors (anonymized), their posts & domain expert annotated labels. | Suicide Risk | [31] | S |
| *Aladaug* | Aladaug *et. al.*, [3] | Contains 10,785 posts which were randomly selected and 785 were manually annotated as suicidal or non-suicidal | Suicidal Id. | [3, 112] | OR |
| CLPsych | Coppersmith *et. al.*, [19] | Depression-v-Control [DvC], PTSD-v-Control [PvC], and Depression-v-PTSD [DvP] | Suicide Risk | [19, 45, 60, 65, 76, 76, 83, 88], | S |
| UMD-RD[9] | Shin *et. al.*, [104] | Contains 11,129 users who posted on r/SuicideWatch and 11,129 users who did not | Suicidal Id. | [31, 60, 104] | S |
| eRISK | Losada *et. al.*, [54] | early risk detection by CLEF lab about problems of detecting depression, anorexia and self-harm | Depression | [4, 54, 102] | ✓ |
| Dreaddit | Turcan *et. al.*, [115] | Consists of 190K Reddit posts of 5 different categories | Stress | [41, 116] | ✓ |
| GoEmotion | Demszky *et. al.*, [24] | manually annotated 58k Reddit comments for 27 emotion categories | Emotion | [116] | ✓ |
| *Pirina* | Pirina *et. al.*, [80] | Given the filtered data from Reddit social medida platform for depression detection task | Depression | [80, 111] | ✓ |
| Sina Weibo | Cao *et. al.*, [10] | Dataset with 3,652 (3,677) users with (without) suicide risk from Sina Weibo | Suicidal Id. | [9, 10] | OR |
| *Ji* | Ji *et. al.*, [46] | Reddit dataset: 5,326 suicidal id. samples out of 20k while Twitter dataset has 594 tweets out of 10k | Suicidal id. | [101, 112] | OR |

Table 7. Results obtained for Social Media Health Detection. ✓: Available, S: Available via Signed agreement, OR: Available on request to authors

*3.1.5 Suicide Risk Assessment using Reddit (SRAR):.* The SRAR dataset[14] is available in public domain. The dataset is composed of 500 Redditors (anonymized), their posts and domain expert annotated labels. The SRAR is used along with different lexicons which are built from the knowledge base associated with mental health like SNOMED-CT, ICD-10, UMLS, and Clinical Trials. This dataset is recently used [31] and the research community is looking forward to use this in near future to enhance the proposed techniques.

*3.1.6 Aladaug:* This dataset is built by Aladaug [3] during his study on suicidal tendency identification from the posts over social media data. Since, there is no name given to this dataset, this dataset is named as *Aladaug* to refer it in this study. Among 10,785 posts, 785 were manually labelled for this study. This dataset is available on request from authors.

*3.1.7 CLPsych 2015 Shared task dataset:* The CLPsych dataset [15] contains three modules which are available via signed agreement, namely, DepressionvControl [DvC], PTSDvControl [PvC], and DepressionvPTSD [DvP]. While distribution of this dataset, the researchers are asked to sign a confidentiality agreement to ensure the privacy of the data.

*3.1.8 The University of Maryland Reddit Suicidality Dataset (UMD-RD).* : The UMD- Reddit Dataset [16] contains one sub-directory with data pertaining to 11,129 users who posted on *SuicideWatch*, and another for 11,129 users who did not. For each user there is full longitudinal data from the 2015 Full Reddit Submission Corpus. The UMD- reddit dataset have been used by academic researchers actively since 2019 as it is available via signed agreement.

*3.1.9 eRISK.* : The eRISK dataset [17] is available online for experiments and analysis to meet the targets of a shared task since few years. The dataset for early risk detection by CLEF lab is given to solve the problems of detecting depression, anorexia and self-harm since few years.

*3.1.10 Dreaddit:* : Dreaddit [18], a new text corpus of lengthy multi-domain social media data for the identification of stress. This dataset consists of 190K posts from five different categories of Reddit communities; the authors additionally label 3.5K total segments taken from 3K posts using Amazon Mechanical Turk. The lexical features which used in this dataset are Dictionary of Affect in Language [123], LIWC features [79] and patterns sentiment library [23]; syntactic features like unigrams and bigrams, the Flesch-Kincaid Grade level and the automated reliability index; social media features like timestamp, upvote ratio, karma (upvote - downvote) and the total number of comments.

*3.1.11 GoEmotion:* The GoEmotion dataset [19] contains 58k carefully curated comments extracted from Reddit, with human annotations to 27 emotion categories or Neutral. It also contains a filtered version based on reter-agreement, which contains a train/test/validation split. This dataset is proposed [24] in 2020 for emotion detection and is used to validate the scalability of the proposed models for stress detection.

*3.1.12 Pirina:* A new dataset is proposed [80], named as Pirina to refer it in this study and is available online [20] for research purposes. A filtered data is extracted from Reddit social media platform for depression detection task. Although, this dataset is not actively maintained, it can be extracted and can be used for pilot study.

[14]https://github.com/AmanuelF/Suicide-Risk-Assessment-using-Reddit
[15]https://github.com/clpsych/shared_task
[16]http://users.umiacs.umd.edu/ resnik/umd_reddit_suicidality_dataset.html
[17]https://erisk.irlab.org/eRisk2021.html
[18]http://www.cs.columbia.edu/~eturcan/data/dreaddit.zip
[19]https://github.com/google-research/google-research/tree/master/goemotions
[20]https://files.pushshift.io/reddit/submissions/

*3.1.13 Sina Weibo:* Another dataset which is proposed for public domain and remains un-named is given the name of the social media platform, Sina Weibo [21], to refer it for this study. The dataset with 3652 users having suicidal tendency and 3677 users not having suicidal risk is extracted from Sina Weibo, a Chinese social media platform.

*3.1.14 Ji:* A new reddit dataset of 5326 suicidal posts out of 20,000 posts were extracted and 594 Suicidal Tweets out of 10,000 Tweets were extracted for experiments and evaluation of the proposed classification approach for suicidal risk detection. This dataset is referred as *Ji* dataset [22] in this study which is available on-request.

### 3.2 The Historical Evolution of Classification Models

In this Section, the evolution of classification techniques are discussed for identifying suicidal tendency of the user on social media. The academic researchers from Microsoft, one of the leading IT based solution organization, have discussed the role of social media in identifying mental health problems. After comprehensive study of 77 research articles on three of the most widely studied mental health problems, stress, depression and suicide risk; the evolution of historical timeline is represented in Fig. 7. The overview of the path breaking baseline classification models for mental health detection are shown in Fig. 8.

The research domain of MIDAS has evolved since 2013 with the path-breaking work on investigating the significance of social media data for predicting the depression [21] and suicidal tendency[59] among users. With introduction to the word embedding and converting information into the vector space [64], there are studies over the psychological stress detection on social media using deep neural network [51, 52]. The academic researchers have explored some more features other than the linguistic features and worked with the user-profile features [83] for MIDAS. The enhancement of this research domain took place after introduction of the available dataset in public domain. It is observed that since many ethical issues are associated with the sensitivity of the data, the release of CLPsych shared task [19] data paves a concrete path for research and development. In this context, the ethical implications of social media data [18] is the benchmark study of dataset constraints for MIDAS.

To extend the studies over different social media platforms, the study for different social media platforms like Facebook [118], Sina weibo (a Chinese online social platform) , and Instagram [14, 85] is observed during 2017. In 2017, the machine learning algorithms were experimented and evaluated for predicting the mental health problems [4][23]. Another important contribution during this phase was the use of social media and social network features for stress detection [53]. Simultaneously, the dual attention mechanism was introduced for multimodal approaches in a study[72]. Thereafter, few studies over the use of attention network [17, 44, 107] across heterogeneous social media data was performed for MIDAS.

The use of deep learning algorithms and supervised approach [46, 76] for MIDAS increased after. More studies revolve around the dimensionality reduction or optimizing the feature vector [76] for machine learning and deep learning learning models, respectively. Few studies handle another challenge of imbalanced dataset for MIDAS [17]. The studies for depression detection started with the use of different social network features [21], evolved with interactions over social media [53] and cascading social networks [67] to extract reliable features, followed by ontology and knowledge graphs [9].

The observations about dynamic Tweets timeline were made while proposing an interpretive Multi-Modal Depression Detection with Hierarchical Attention Network (MDHAN) [135]. The MDHAN framework was designed by using multi-model features and two attention mechanisms are applied at tweet-level and at word-level. Further, to deal with some multimodal features, another depression detection mechanism was identified as COMMA [36]. The idea is to explore more features which include both textual and visual understanding of the data. Some recent multimodal feature extraction techniques are multiple social networking learning (MSNL) [108] , Wasserstein dictionary learning (WDL) [89], and multimodal depressive dictionary learning (MDL) [102] methods. The authors in Dual-ContextBERT model [60] have used the multi-level model by removing the limitation of single level analysis.

[21]https://github.com/bryant03/Sina-Weibo-Dataset
[22]https://github.com/shaoxiongji/sw-detection
[23]https://github.com/BigMiners/eRisk2017

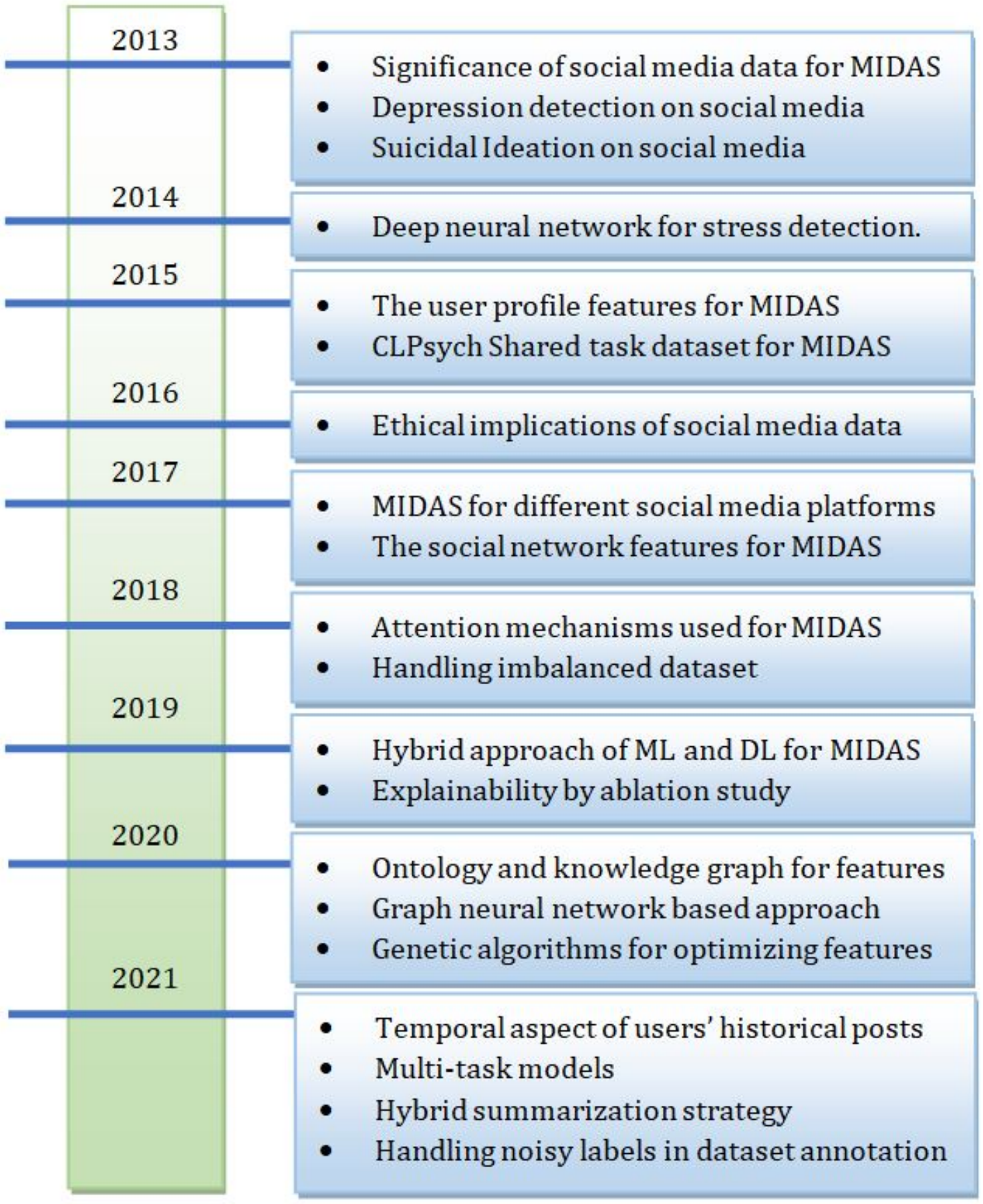

Fig. 7. The timeline of evolving important events for quantification of suicidal tendency on social media

The responsible and explainable models [93, 116] for mental health prediction by using the *Ablation study*. After extensive literature over Machine Learning (ML) and Deep Learning (DL) algorithms for classification problems, few studies were conducted with hybridization of ML and DL approaches [60]. Recently, few other studies have chosen to investigate the Graph Neural Network for MIDAS [9]. To further improve the feature optimization, the computation intelligence techniques like genetic algorithm [101] were used for MIDAS. Recently, the ordinal attention network was proposed [94][24] for suicidal ideation detection

[24]https://github.com/midas-research/sismo-wsdm

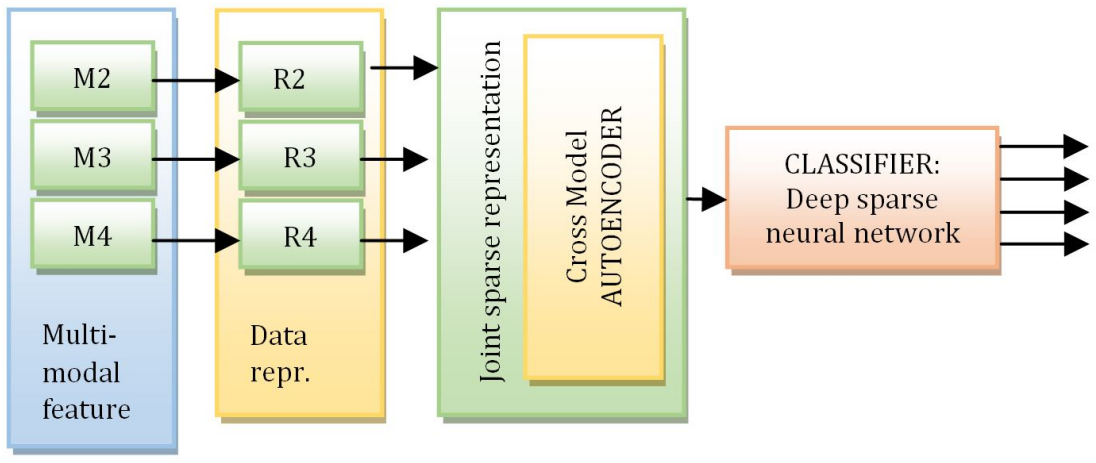

(a) CAE: Cross-model Auto-Encoder

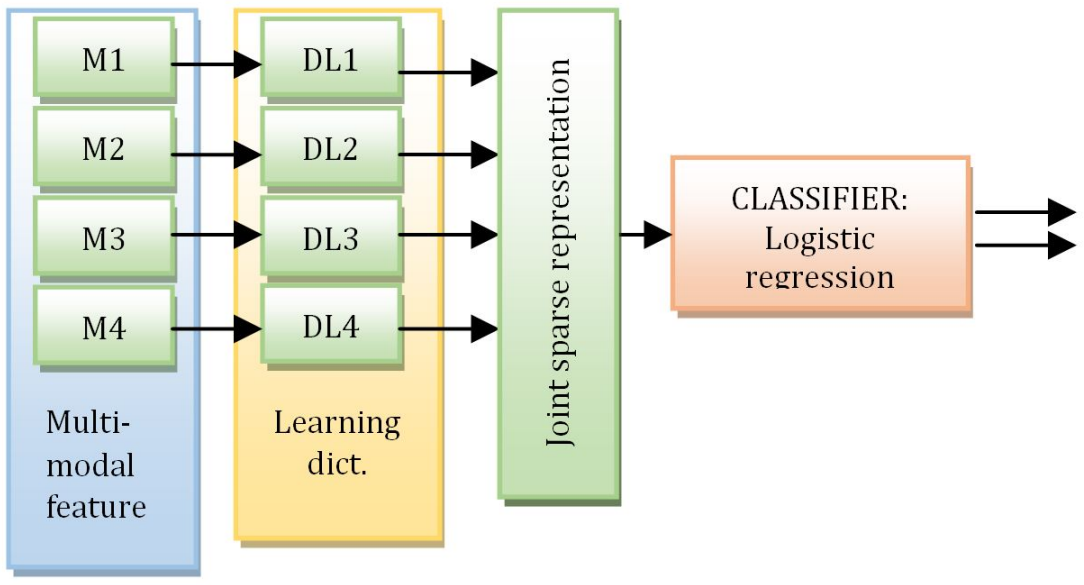

(b) MDL: Multimodal Dictionary Learning

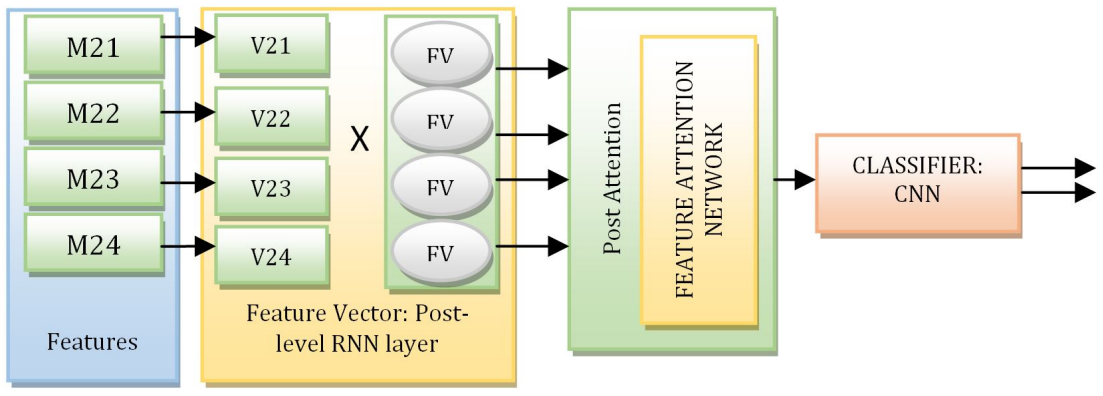

(c) Feature Attention Network

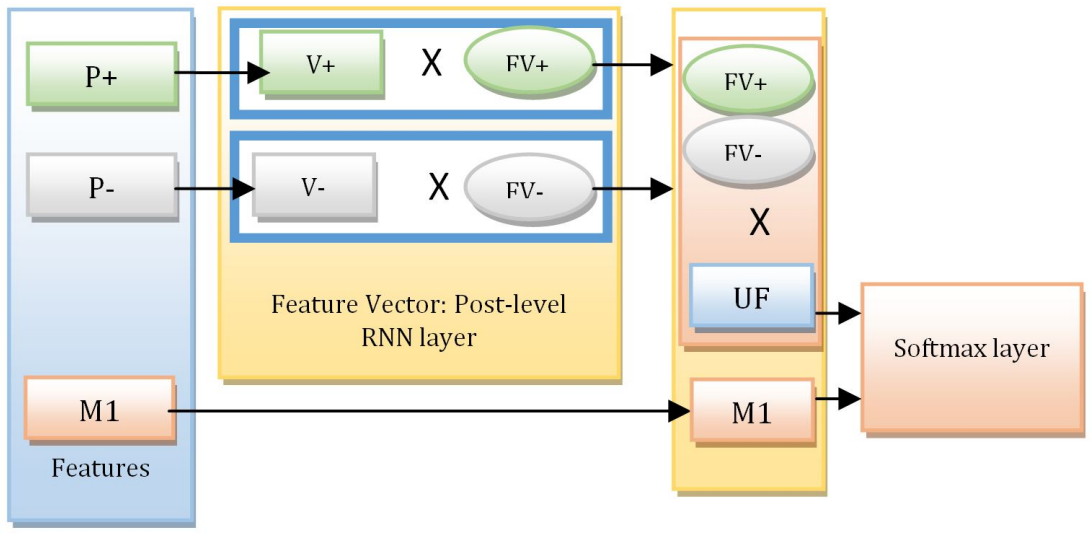

(d) DualContextBERT for processing features

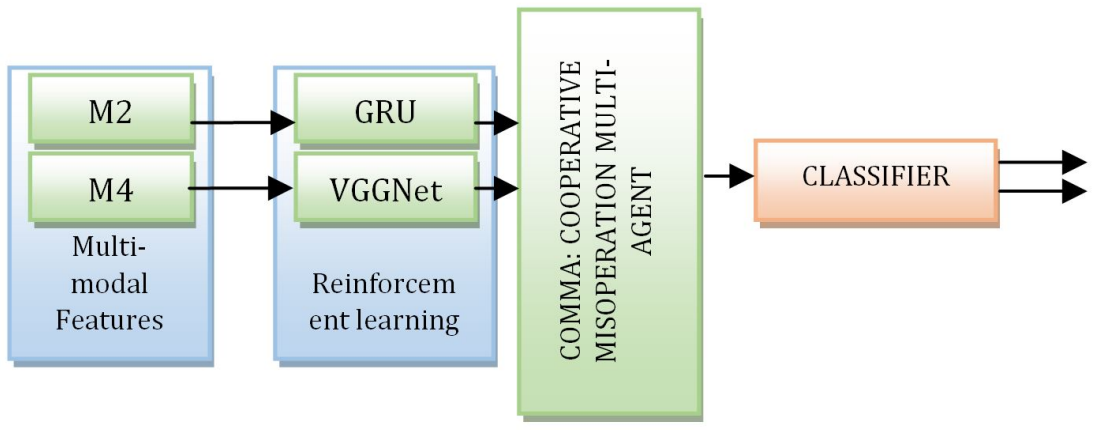

(e) COMMA: COoperative Misoperation Multi-Agent

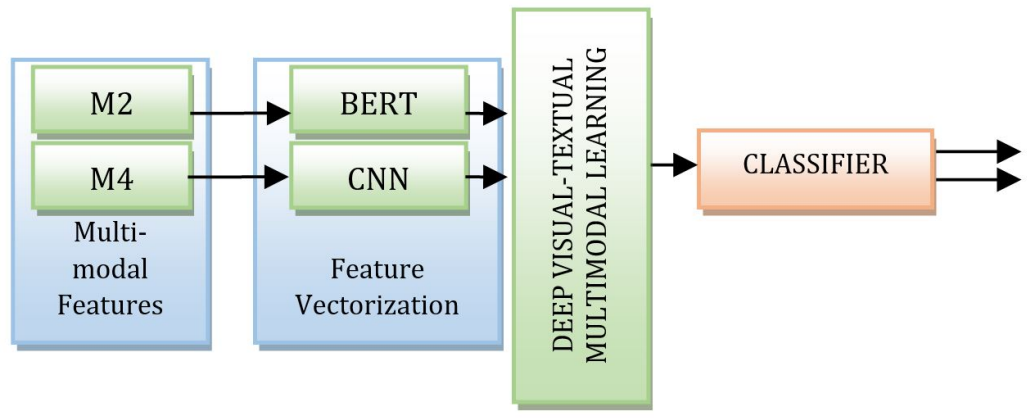

(f) SenseMood: A deep visual-textual multimodal learning

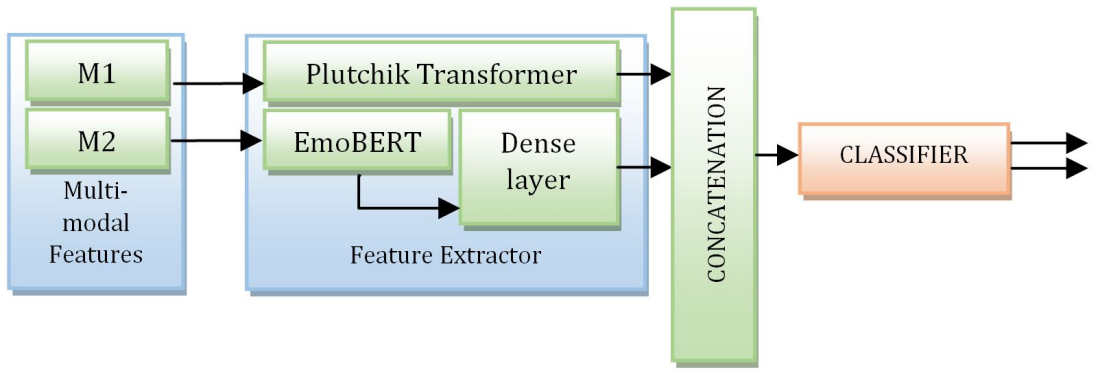

(g) PHASE: Phase adaptive convolution for users' timeline

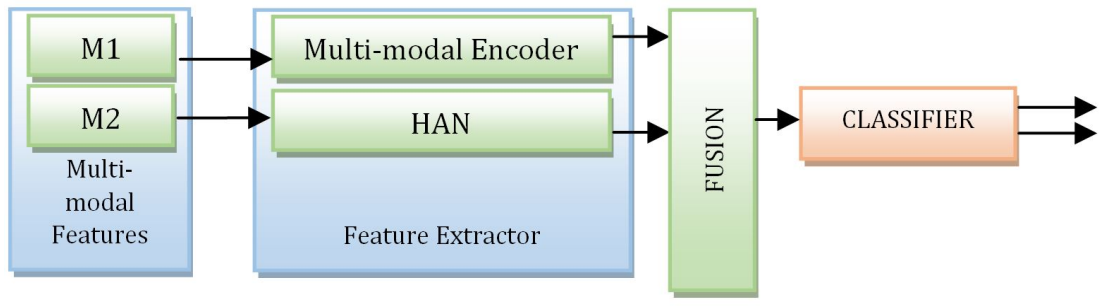

(h) MDHAN: Multimodal Depression detection HAN

Fig. 8. Some existing models for quantifying the suicidal tendency on social media

The recent research areas of MIDAS comprise of many different dimensions, for instance, the noisy label detection for automatic annotation of dataset [40]; the historical aspect of the users' timeline for identifying different phase of mental health [93], the hybrid extractive and abstractive summarization strategy as DepressionNet [133][25]. The multitask models [116][26] were observed for emotions, stress and depression detection.

[25]https://github.com/hzogan/DepressionNet
[26]https://github.com/eturcan/emotion-infused

| Outcome | Paper | Method | Dataset | Baselines | Results | C. |
|---|---|---|---|---|---|---|
| Depression | Shen *et. al.,* [102] | MDL | MDDL | Naive Bayes [89, 108] | F1: 85% | |
| Suicidal Id. | Sawhney *et. al.,* [93] | PHASE | Self[96] | [10, 31, 60, 96, 97, 106] | F1: 80.5% | ✓ |
| Depression | Gui *et. al.,* [36] | COMMA | MDDL | Naive Bayes, [89, 102, 108] | F1: 90% | |
| Stress | Lin *et. al.,* [51] | CAE | Self | (SVM, ANN, DNN) + SAE | F1: 86.12% | |
| Stress | Lin *et. al.,* [53] | FGM | Self | LR, SVM, RF, GBDT, DNN | F1: 93.40% | |
| Depression | Song *et. al.,* [107] | FAN+CNN | RSDD | RNN | Better P | |
| Suicidal Id. | Sawhney *et. al.,* [96] | C-LSTM | Self[96] | LSTM, RNN | F1: 82.7% | |
| Suicidal Id. | Sawhney *et. al.,* [94] | SISMO | SRAR | [5, 10, 31, 60, 97] | F1: 73% | |
| Suicidal Id. | Matero *et. al.,* [60] | D-C BERT | Self | BERT, Dict. | F1: 50% | |
| Depression | Orabi *et. al.,* [76] | WEO | CLPsych | Word2vec | F1: 86.96% | |
| Dep./ SI | Haque *et. al.,* [40] | GUSE | SDCNL | BERT | F1: 95.44% | ✓ |
| Depression | Almeida *et. al.,* [4] | ML | eRISK | ML | F1: 53% | ✓ |
| Depression | Zogan *et. al.,* [135] | MDHAN | MDDL | HAN, CNN, [102], BiGRU | F1: 89.3% | |
| Stress | Turcan *et. al.,* [116] | *Turcan* | Dreaddit+1 | RNN, BERT, Multi-task | F1: 80.34% | ✓ |
| Depression | Lin *et. al.,* [50] | *Lin* | MDDL | [69, 72, 102, 103] | F1: 93.60% | |
| Depression | Tadesse *et. al.,* [111] | MLP | Pirina | SVM, LR, RF, Ada-Boost | F1: 93% | |
| Depression | Cong *et. al.,* [17] | XA-Boost | RSDD | SVM, LSTM, [79, 129] | F1: 60% | |
| Depression | Zogan *et. al.,* [133] | DepressionNet | MDDL | [15, 102], BiGRU, CNN | F1:91.2% | ✓ |
| Suicidal Id. | Mishra *et. al.,* [67] | snap-batnet | Self[96] | [96], ELMo, RCNN | F1:92.6% | |
| Suicidal Id. | Tadesse *et. al.,* [112] | LSTM-CNN | Self | RF, SVM, NB, XGBOOST | F1: 93.4% | ✓ |
| Suicidal Id. | Cao *et. al.,* [10] | SDM | Sina-W | NB, LSTM, SVM | F1: 90.92% | |
| Suicidal Id. | Cao *et. al.,* [9] | KGbased | Sina-W | SDM, LSTM, CNN | F1: 93.69% | |
| Suicidal Id. | Ma *et. al.,* [58] | DAM | Self | NB, LSTM, SVM, CNN [10] | F1: 91.54% | |

Table 8. Linguistic feature extraction for mental health status.

Few existing studies on the multi-modal feature extraction techniques is used for depression detection [36] like Co-attention [55], Dual attention[72], and Modality attention [69]. The novel contributions for suicidal tendency predictions are comprised of new feature harvesting [93, 102], feature tranformation or transformation learning [51, 107], classfication techniques [36], and interpretability and explainability [93, 107].

### 3.3 Summary

To summarize the extensive study of classification models for identifying suicidal tendency, the dataset, baselines, results and code availability is shown in Table 8. The classification techniques for identification of stress, depression and suicide risk are not directly comparable because there is no benchmark dataset which is available in public domain (due to sensitivity of the dataset). However, while indirectly comparing the proposed techniques, it is observed that SDCNL[27] gives the best performance while classifying depression and suicidal posts by utilizing the transformer for data representation. The LSTM-CNN[28] [112] and KGbased [9] approach have shown improved performance over self defined *English* dataset and Sina Weibo *Chinese* dataset, respectively.

It is interesting to note that many other techniques have been proposed for stress and depression which can be used for parallel and sequential neural networks for experiments and analysis over suicidal risk detection in near future.

[27] https://github.com/ayaanzhaque/SDCNL
[28] https://github.com/shaoxiongji/sw-detection

Before the detailed discussion over inferences on existing studies, the basic tools and resources which are used for this study are discussed in this Section.

### 3.4 Tools and resources

As discussed earlier, the social media data is first-hand user-generated information which may or may not be ill-formed due to many spelling mistakes, use of abbreviations and slang. To identify named entities and semantics among them is still a challenging task. The NLTK library of Python is used for handling ill-formed text. In addition to the classical tools and libraries for text classification, few other libraries are used for this task of suicidal tendency detection on social media.

*3.4.1 Python Reddit API.* The Reddit social media platform is the most widely used resource for identifying users with stress, depression and suicidal risk. The Python Reddit API Wrapper (PRAW)[29] aims to scrap the posts from Reddit for detection and analysis. The PRAW API follows all the Reddit API rules [30] for scrapping posts from Reddit.

*3.4.2 PyPlutchik.* The embedding is fine tuned in some studies using the existing emotion models of Plutchik and available as pre-built tools in Python environment. The most recently used such tools are the PyPlutchik tool [100] which was built from Plutchik model of emotions as introduced in PHASE [93]suicidal ideation detection.

*3.4.3 BERT Summarizer.* The BERT summarizer[31] is used in existing literature along with DistilBERT [32] for social mental health detection. In this context, few other feature transformation models are used for feature vectorization by using *HuggingFace* library [33].

*3.4.4 DLATK Python Package.* DLATK stands for Differential Language Analysis Toolkit which is an end to end human text analysis package, specifically suited for social media and social scientific applications. All non-neural models were implemented via the DLATK Python package [99] as used for social mental health detection while introducing Dual-Context BERT [37, 60].

*3.4.5 Application with Facebook.* Adaptive experimentation is the machine-learning guided process of iteratively exploring a (possibly infinite) parameter space in order to identify optimal configurations in a resource-efficient manner. Ax[34] currently supports Bayesian optimization and bandit optimization as exploration strategies. Bayesian optimization in Ax is powered by BoTorch, a modern library for Bayesian optimization research built on PyTorch. Ax is used for Social mental health detection [116].

## 4 DISCUSSION

It is observed that the best approach for depression detection is recently proposed as SenseMood in 2020 [50] and validated over MDDL dataset. The SenseMood outperforms more than ten baselines directly and indirectly. After extensive study of 77 research articles related to stress, depression and suicidal tendency, various inferences are made in existing study as shown in Table 10 and Table 11, and new research directions as shown in Fig. 9 are discussed in this Section.

[29]https://github.com/praw-dev/praw, https://github.com/shaoxiongji/webspider
[30]https://github.com/reddit-archive/reddit/wiki/API
[31]https://github.com/dmmiller612/bert-extractive-summarizer
[32]https://huggingface.co/transformers/model_doc/distilbert.html
[33]https://huggingface.co/
[34]https://github.com/facebook/Ax

| **Paper** | **Year** | **Contributions** | **Det.** | **BA** | **Str** | **Ex** | **LI** | **Code** | **L** |
|---|---|---|---|---|---|---|---|---|---|
| Lin *et. al.,* [51] | 2014 | proposed a cross-media auto-encoder for joint representation of features | ✓ | | | | | R | Chinese |
| Lin *et. al.,* [53] | 2017 | proposed factor graph model (FGM) with CNN for classification | ✓ | ✓ | | | ✓ | R | Chinese, English |
| Shen *et. al.,* [102] | 2017 | Public dataset, feature extraction with scalable approach for SMHP | ✓ | ✓ | | | | R | English |
| Almeida *et. al.,* [4] | 2017 | Checked different machine learning models with classification. | ✓ | | | | | A | English |
| Song *et. al.,* [107] | 2018 | Proposed a feature attention network for identifying important features | ✓ | | | I | | R | English |
| Orabi *et. al.,* [76] | 2018 | Proposed the Word Embedding Optimization (WEO) for optimizing the feature vectors | ✓ | | | | | NA | English |
| Gui *et. al.,* [36] | 2019 | Proposed GRU + VGGNet + COMMA model for Depression detection | ✓ | ✓ | | | | R | English |
| Matero *et. al.,* [60] | 2019 | proposed a dual context based approach by hybridising both ML and DL | ✓ | | | | | NA | English |
| Guntuku *et. al.,* [37] | 2019 | Implications of using social media as a tool for stress detection, studies over Facebook and Twitter | | ✓ | | | | NA | English |
| De Choudhury *et. al.,* [21] | 2013 | Work on statistics of social media data which can be used for SMHP. | | ✓ | | | | NA | English |
| Tadesse *et. al.,* [111] | 2019 | Investigated the machine learning techiques for depression detection | ✓ | | | | | R | English |
| Cong *et. al.,* [17] | 2018 | proposed the model by integrating XGBoost and Attention with BiLSTM | ✓ | | | | | R | English |
| Mishra *et.al.,* [67] | 2019 | proposed the social networking features based model for identifying suicide ideation | ✓ | | | | | R | English |
| Cao *et. al.,* [10] | 2019 | proposed a new model SDM with two-layered attention mechanism and domain specific word embeddings | ✓ | ✓ | | ✓ | ✓ | R | Chinese |
| Vioules *et. al.,* [119] | 2018 | automatic identification of user's online behaviour | ✓ | ✓ | ✓ | | | NA | English |

Table 10. Inferences of evolving suicidal tendency detection on social media. Ex: Explainability, A: Available, R: Reproducible, S: Available by Signed Agreement, NA: Not Available, Str: Streming Data, LI: Language Independent, L: Language used, Det.: Detection, BA: Behavioural Analysis

| Paper | Year | Contributions | Det. | BA | Str | Ex | LI | Code | L |
|---|---|---|---|---|---|---|---|---|---|
| Xu *et. al.,* [127] | 2020 | Jointly analyzing language, visual, and metadata cues and their relation to mental health | ✓ | | | | | R | English |
| Lin *et. al.,* [50] | 2020 | Proposed deep visual textual multimodal learning to map psychological state of users on social media | ✓ | ✓ | ✓ | | | R | English |
| Tadesse *et. al.,* [112] | 2020 | proposed the LSTM + CNN classification model | ✓ | | | | | R | English |
| Cao *et. al.,* [9] | 2020 | proposed a knowledge graph and ontology based graphical neural network for suicide risk detection | ✓ | ✓ | | ✓ | ✓ | R | Chinese, English |
| Shah *et. al.,* [101] | 2020 | Proposed hybrid approach by using computationally intelligent techniques and other optimizations for features | ✓ | | | | | R | English |
| Sawhney *et. al.,* [93] | 2021 | Users' historical timeline encoded and mapped with other features | ✓ | | | ✓ | | A | English |
| Sawhney *et. al.,* [94] | 2021 | Proposed an ordinal attention network for suicidal ideation detection | ✓ | ✓ | | ✓ | | A | English |
| Zogan *et. al.,* [135] | 2021 | Proposed multi-modal depression detection with HAN (MDHAN) | ✓ | | | ✓ | | R | English |
| Turcan *et. al.,* [116] | 2021 | Multi-task with emotional models for more explainable stress detection model | ✓ | ✓ | | ✓ | | A | English |
| Haque *et. al.,* [40] | 2021 | Proposed SDCNL model with GUSE - dense over UMAP- (Kmeans, GMM) | ✓ | | | | | A | English |
| Zogan *et. al.,* [133] | 2021 | proposed a DepressionNet using hybrid extractive and abstractive summarization strategy | ✓ | ✓ | | | | A | English |

Table 11. Inferences of recent suicidal tendency detection on social media. Ex: Explainability, A: Available, R: Reproducible, S: Available by Signed Agreement, NA: Not Available, Str: Streming Data, LI: Language Independent, L: Language used, Det.: Detection, BA: Behavioural Analysis

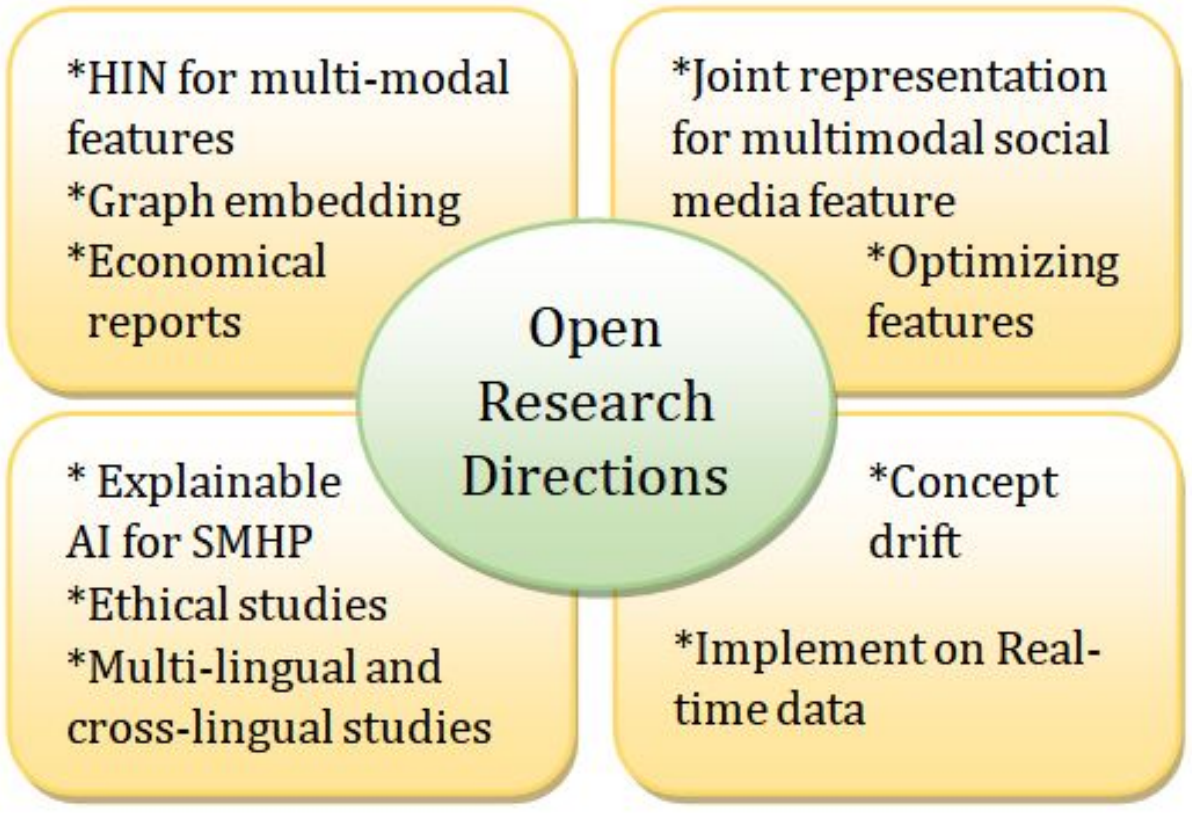

Fig. 9. Open challenges and new research directions in identifying suicidal tendency on social media

### 4.1 Noisy Labels

Recently, the problem of noisy labels is identified in existing studies where the potential of some labels of the data is found to be corrupted and hence, is mentioned as the noisy labels. To solve this problem, SDCNL model has introduced a unique feature of label correction methodology [40] for classifying posts as suicide versus depression.

### 4.2 Feature Extraction

The other factors which can be potential features are the happiness index of the country of a user; the ease of living index of the country of the user; the variation in geographical locations and multi-source distributed crawling; detection of multi sources communities by using spectral clustering over multi-level graphs [28]. Although there are studies on finding correlations among different features and map the variables for mental illness health detection in China [49], there is the need to study this for different countries and at global level due to much of socio-political difference in each country.

### 4.3 Embedding for Multi-task problem

Recently, many academic researchers are using the embedding average for using word embedding across multi-task problems, however, a systematic word embedding optimizer is introduced for multi-task problem of mental health prediction [76]. However, there is no explainability or mathematical validation for why the results are better.

### 4.4 Time Complexity

Although, it is observed that the recent approaches for stress detection shows the significant improvement with F1-score for FGM as 93.40 [52], however, it is computationally expensive and takes almost more than the double time as compared to the second best approach. There is need to reduce the complexity of the proposed techniques. There are few more studies analyzing the time complexity of the mental health detection techniques [6].

### 4.5 Behavioral Analysis

The mental health detection is the part of integrated study of computational linguistics, human-computer interactions and clinical psychology. Few studies have observed the latent patterns among social media users to express their sensitive thoughts are found to be common. Authors argue that depressed tweets are more likely to be expressed late night [102] than during day time. This analytical part of human behavior is rarely explored in the existing literature as observed from Table 10 and Table 11.

### 4.6 Interpretability and Explainability

There are detailed and theoretical explanations of the proposed approach to test its interpretability [107] or explainability[9, 93, 116, 135]. The ablation studies[9, 93] and LIME[116] study are observed in literature and this section of ethical validation can be explored further to enhance the applicability of the proposed method in real-time applications.

### 4.7 Social Networks and Graph Neural Networks

It is observed that many different modalities are being used for social mental health analysis. The trend of utilizing textual, visual, and multimedia information has given a new research direction in this domain. Although, few studies have exploit the network features for Twitter data [130], still there is a big room to study multi-level networks and heterogeneous information networks for multi-modal information in social media for better and integrated representation. Few studies on knowledge graph [9], ontology [9] and graph neural networks [95] validate it as a progressive domain.

### 4.8 Multi-lingual, cross-lingual and language-independent approach

It is assumed that unless mentioned, the existing studies have not used any other language specific measures except English. Authors have given new research direction to multi-task learning [116, 127] but there is no work found in the multi-lingual approach as observed for offensive language [84]. Few studies have made progress towards language independent approach [9, 10, 53], however, the existing techniques are not directly or indirectly not compared for language-independent or multi-lingual approach.

### 4.9 Incremental Learning from Streaming Data

There are some studies on Topic extraction on social media content for early depression detection on retrospective data [61] and phase change of the user [93, 95]. However, the existing studies have rarely used the online streaming data [50] and there is no such study which shows the concept drift [119] in streaming data. This concept drift may help in identifying the level of changing risk in suicidal tendency.

### 4.10 Real-time Applications

The real-time mental health prediction is yet to be explored as minimal studies are observed on integration of Internet of Medical Things (IoMT) and Social Media dataset by academic researchers [39].

## 5 CONCLUSION

This manuscript can be summarized as an extensive literature survey on predicting the suicidal tendency from social media data. The exponential progress in the field of data science for mental health prediction has shown its significance in recent years. The corpus of 77 research articles contains studies over stress, depression and suicide risk detection on social media. However, there is no substantial work on quantifying the suicidal risk from the longitudnal data of the user. To handle this and to integrate the exsiting studies on multiple tasks, an extensive survey is given along with the open challenges and possible research directions. The major contributions of this manuscript are enlisting the available dataset (publicly, on-request and via signed agreement); introduction to the taxonomy of the mental healthcare; classification of feature extraction and transformation techniques for vector representation; the historical evolution of suicidal tendency detection with timeline; new research directions and open challenges. This manuscript further highlights the important contributions which can be used as benchmark studies in this domain.

## REFERENCES

[1] Muhammad Abdul-Mageed and Lyle Ungar. 2017. Emonet: Fine-grained emotion detection with gated recurrent neural networks. In *Proceedings of the 55th annual meeting of the association for computational linguistics (volume 1: Long papers)*. 718–728.

[2] Nafiz Al Asad, Md Appel Mahmud Pranto, Sadia Afreen, and Md Maynul Islam. 2019. Depression detection by analyzing social media posts of user. In *2019 IEEE International Conference on Signal Processing, Information, Communication & Systems (SPICSCON)*. IEEE, 13–17.

[3] Ahmet Emre Aladağ, Serra Muderrisoglu, Naz Berfu Akbas, Oguzhan Zahmacioglu, and Haluk O Bingol. 2018. Detecting suicidal ideation on forums: proof-of-concept study. *Journal of medical Internet research* 20, 6 (2018), e215.

[4] Hayda Almeida, Antoine Briand, and Marie-Jean Meurs. 2017. Detecting Early Risk of Depression from Social Media User-generated Content.. In *CLEF (Working Notes)*.

[5] Payam Amini, Hasan Ahmadinia, Jalal Poorolajal, and Mohammad Moqaddasi Amiri. 2016. Evaluating the high risk groups for suicide: a comparison of logistic regression, support vector machine, decision tree and artificial neural network. *Iranian journal of public health* 45, 9 (2016), 1179.

[6] Sergio G Burdisso, Marcelo Errecalde, and Manuel Montes-y Gómez. 2019. A text classification framework for simple and effective early depression detection over social media streams. *Expert Systems with Applications* 133 (2019), 182–197.

[7] Rafael A Calvo, David N Milne, M Sazzad Hussain, and Helen Christensen. 2017. Natural language processing in mental health applications using non-clinical texts. *Natural Language Engineering* 23, 5 (2017), 649–685.

[8] Erik Cambria, Daniel Olsher, and Dheeraj Rajagopal. 2014. SenticNet 3: a common and common-sense knowledge base for cognition-driven sentiment analysis. In *Proceedings of the AAAI Conference on Artificial Intelligence*, Vol. 28.

[9] Lei Cao, Huijun Zhang, and Ling Feng. 2020. Building and Using Personal Knowledge Graph to Improve Suicidal Ideation Detection on Social Media. *IEEE Transactions on Multimedia* (2020).

[10] Lei Cao, Huijun Zhang, Ling Feng, Zihan Wei, Xin Wang, Ningyun Li, and Xiaohao He. 2019. Latent suicide risk detection on microblog via suicide-oriented word embeddings and layered attention. *arXiv preprint arXiv:1910.12038* (2019).

[11] Daniel Cer, Yinfei Yang, Sheng-yi Kong, Nan Hua, Nicole Limtiaco, Rhomni St John, Noah Constant, Mario Guajardo-Céspedes, Steve Yuan, Chris Tar, et al. 2018. Universal sentence encoder. *arXiv preprint arXiv:1803.11175* (2018).

[12] Stevie Chancellor, Michael L Birnbaum, Eric D Caine, Vincent MB Silenzio, and Munmun De Choudhury. 2019. A taxonomy of ethical tensions in inferring mental health states from social media. In *Proceedings of the conference on fairness, accountability, and transparency*. 79–88.

[13] Stevie Chancellor and Munmun De Choudhury. 2020. Methods in predictive techniques for mental health status on social media: a critical review. *NPJ digital medicine* 3, 1 (2020), 1–11.

[14] Qijin Cheng, Tim MH Li, Chi-Leung Kwok, Tingshao Zhu, and Paul SF Yip. 2017. Assessing suicide risk and emotional distress in Chinese social media: a text mining and machine learning study. *Journal of medical internet research* 19, 7 (2017), e243.

[15] Junyoung Chung, Caglar Gulcehre, KyungHyun Cho, and Yoshua Bengio. 2014. Empirical evaluation of gated recurrent neural networks on sequence modeling. *arXiv preprint arXiv:1412.3555* (2014).

[16] Arman Cohan, Bart Desmet, Andrew Yates, Luca Soldaini, Sean MacAvaney, and Nazli Goharian. 2018. SMHD: a large-scale resource for exploring online language usage for multiple mental health conditions. *arXiv preprint arXiv:1806.05258* (2018).

[17] Qing Cong, Zhiyong Feng, Fang Li, Yang Xiang, Guozheng Rao, and Cui Tao. 2018. XA-BiLSTM: A deep learning approach for depression detection in imbalanced data. In *2018 IEEE International Conference on Bioinformatics and Biomedicine (BIBM)*. IEEE, 1624–1627.

[18] Mike Conway and Daniel O'Connor. 2016. Social media, big data, and mental health: current advances and ethical implications. *Current opinion in psychology* 9 (2016), 77–82.

[19] Glen Coppersmith, Mark Dredze, Craig Harman, Kristy Hollingshead, and Margaret Mitchell. 2015. CLPsych 2015 shared task: Depression and PTSD on Twitter. In *Proceedings of the 2nd Workshop on Computational Linguistics and Clinical Psychology: From Linguistic Signal to Clinical Reality*. 31–39.

[20] Munmun De Choudhury. 2013. Role of social media in tackling challenges in mental health. In *Proceedings of the 2nd international workshop on Socially-aware multimedia*. 49–52.

[21] Munmun De Choudhury, Michael Gamon, Scott Counts, and Eric Horvitz. 2013. Predicting depression via social media. In *Proceedings of the International AAAI Conference on Web and Social Media*, Vol. 7.

[22] Munmun De Choudhury, Emre Kiciman, Mark Dredze, Glen Coppersmith, and Mrinal Kumar. 2016. Discovering shifts to suicidal ideation from mental health content in social media. In *Proceedings of the 2016 CHI conference on human factors in computing systems*. 2098–2110.

[23] Tom De Smedt and Walter Daelemans. 2012. Pattern for python. *The Journal of Machine Learning Research* 13, 1 (2012), 2063–2067.

[24] Dorottya Demszky, Dana Movshovitz-Attias, Jeongwoo Ko, Alan Cowen, Gaurav Nemade, and Sujith Ravi. 2020. Goemotions: A dataset of fine-grained emotions. *arXiv preprint arXiv:2005.00547* (2020).

[25] Jacob Devlin, Ming-Wei Chang, Kenton Lee, and Kristina Toutanova. 2018. Bert: Pre-training of deep bidirectional transformers for language understanding. *arXiv preprint arXiv:1810.04805* (2018).

[26] Johannes C Eichstaedt, Robert J Smith, Raina M Merchant, Lyle H Ungar, Patrick Crutchley, Daniel Preoţiuc-Pietro, David A Asch, and H Andrew Schwartz. 2018. Facebook language predicts depression in medical records. *Proceedings of the National Academy of Sciences* 115, 44 (2018), 11203–11208.

[27] Jeffrey L Elman. 1990. Finding structure in time. *Cognitive science* 14, 2 (1990), 179–211.

[28] Aleksandr Farseev, Ivan Samborskii, and Tat-Seng Chua. 2016. A Big Data Platform for Social Multimedia Analytics. (2016).

[29] Elizabeth Ford, Keegan Curlewis, Akkapon Wongkoblap, and Vasa Curcin. 2019. Public opinions on using social media content to identify users with depression and target mental health care advertising: mixed methods survey. *JMIR mental health* 6, 11 (2019), e12942.

[30] Prasadith Kirinde Gamaarachchige and Diana Inkpen. 2019. Multi-task, multi-channel, multi-input learning for mental illness detection using social media text. In *Proceedings of the Tenth International Workshop on Health Text Mining and Information Analysis (LOUHI 2019)*. 54–64.

[31] Manas Gaur, Amanuel Alambo, Joy Prakash Sain, Ugur Kursuncu, Krishnaprasad Thirunarayan, Ramakanth Kavuluru, Amit Sheth, Randy Welton, and Jyotishman Pathak. 2019. Knowledge-aware assessment of severity of suicide risk for early intervention. In *The World Wide Web Conference*. 514–525.

[32] Felipe T Giuntini, Mirela T Cazzolato, Maria de Jesus Dutra dos Reis, Andrew T Campbell, Agma JM Traina, and Jo Ueyama. 2020. A review on recognizing depression in social networks: challenges and opportunities. *Journal of Ambient Intelligence and Humanized Computing* 11, 11 (2020), 4713–4729.

[33] George Gkotsis, Anika Oellrich, Tim Hubbard, Richard Dobson, Maria Liakata, Sumithra Velupillai, and Rina Dutta. 2016. The language of mental health problems in social media. In *Proceedings of the Third Workshop on Computational Linguistics and Clinical Psychology*. 63–73.

[34] Kristina Gligorić, Ashton Anderson, and Robert West. 2018. How constraints affect content: The case of Twitter's switch from 140 to 280 characters. In *Proceedings of the International AAAI Conference on Web and Social Media*, Vol. 12.

[35] Robert N Golden, Carla Weiland, and Fred Peterson. 2009. *The truth about illness and disease*. Infobase Publishing.

[36] Tao Gui, Liang Zhu, Qi Zhang, Minlong Peng, Xu Zhou, Keyu Ding, and Zhigang Chen. 2019. Cooperative multimodal approach to depression detection in Twitter. In *Proceedings of the AAAI Conference on Artificial Intelligence*, Vol. 33. 110–117.

[37] Sharath Chandra Guntuku, Anneke Buffone, Kokil Jaidka, Johannes C Eichstaedt, and Lyle H Ungar. 2019. Understanding and measuring psychological stress using social media. In *Proceedings of the International AAAI Conference on Web and Social Media*, Vol. 13. 214–225.

[38] Sharath Chandra Guntuku, David B Yaden, Margaret L Kern, Lyle H Ungar, and Johannes C Eichstaedt. 2017. Detecting depression and mental illness on social media: an integrative review. *Current Opinion in Behavioral Sciences* 18 (2017), 43–49.

[39] Divya Gupta, MPS Bhatia, and Akshi Kumar. 2021. Real-Time Mental Health Analytics Using IoMT and Social Media Datasets: Research and Challenges. *Available at SSRN 3842818* (2021).

[40] Ayaan Haque, Viraaj Reddi, and Tyler Giallanza. 2021. Deep Learning for Suicide and Depression Identification with Unsupervised Label Correction. *arXiv preprint arXiv:2102.09427* (2021).

[41] Keith Harrigian, Carlos Aguirre, and Mark Dredze. 2020. On the State of Social Media Data for Mental Health Research. *arXiv preprint arXiv:2011.05233* (2020).

[42] Sepp Hochreiter and Jürgen Schmidhuber. 1997. Long short-term memory. *Neural computation* 9, 8 (1997), 1735–1780.

[43] Jamil Hussain, Fahad Ahmed Satti, Muhammad Afzal, Wajahat Ali Khan, Hafiz Syed Muhammad Bilal, Muhammad Zaki Ansaar, Hafiz Farooq Ahmad, Taeho Hur, Jaehun Bang, Jee-In Kim, et al. 2020. Exploring the dominant features of social media for depression detection. *Journal of Information Science* 46, 6 (2020), 739–759.

[44] Julia Ive, George Gkotsis, Rina Dutta, Robert Stewart, and Sumithra Velupillai. 2018. Hierarchical neural model with attention mechanisms for the classification of social media text related to mental health. In *Proceedings of the Fifth Workshop on Computational Linguistics and Clinical Psychology: From Keyboard to Clinic*. 69–77.

[45] Zunaira Jamil. 2017. *Monitoring tweets for depression to detect at-risk users*. Ph.D. Dissertation. Université d'Ottawa/University of Ottawa.

[46] Shaoxiong Ji, Celina Ping Yu, Sai-fu Fung, Shirui Pan, and Guodong Long. 2018. Supervised learning for suicidal ideation detection in online user content. *Complexity* 2018 (2018).

[47] Jia Jia. 2018. Mental Health Computing via Harvesting Social Media Data.. In *IJCAI*. 5677–5681.

[48] Victor Leiva and Ana Freire. 2017. Towards suicide prevention: early detection of depression on social media. In *International Conference on Internet Science*. Springer, 428–436.

[49] He Li, Yujin Han, Yunyu Xiao, Xingyun Liu, Ang Li, and Tingshao Zhu. 2021. Suicidal ideation risk and socio-cultural factors in China: A longitudinal study on social media from 2010 to 2018. *International Journal of Environmental Research and Public Health* 18, 3 (2021), 1098.

[50] Chenhao Lin, Pengwei Hu, Hui Su, Shaochun Li, Jing Mei, Jie Zhou, and Henry Leung. 2020. Sensemood: Depression detection on social media. In *Proceedings of the 2020 International Conference on Multimedia Retrieval*. 407–411.

[51] Huijie Lin, Jia Jia, Quan Guo, Yuanyuan Xue, Jie Huang, Lianhong Cai, and Ling Feng. 2014. Psychological stress detection from cross-media microblog data using deep sparse neural network. In *2014 IEEE International Conference on Multimedia and Expo (ICME)*. IEEE, 1–6.

[52] Huijie Lin, Jia Jia, Quan Guo, Yuanyuan Xue, Qi Li, Jie Huang, Lianhong Cai, and Ling Feng. 2014. User-level psychological stress detection from social media using deep neural network. In *Proceedings of the 22nd ACM international conference on Multimedia*. 507–516.

[53] Huijie Lin, Jia Jia, Jiezhong Qiu, Yongfeng Zhang, Guangyao Shen, Lexing Xie, Jie Tang, Ling Feng, and Tat-Seng Chua. 2017. Detecting stress based on social interactions in social networks. *IEEE Transactions on Knowledge and Data Engineering* 29, 9 (2017), 1820–1833.

[54] David E Losada, Fabio Crestani, and Javier Parapar. 2018. Overview of eRisk: early risk prediction on the internet. In *International conference of the cross-language evaluation forum for european languages*. Springer, 343–361.

[55] Jiasen Lu, Jianwei Yang, Dhruv Batra, and Devi Parikh. 2016. Hierarchical question-image co-attention for visual question answering. *arXiv preprint arXiv:1606.00061* (2016).

[56] Jianhong Luo, Jingcheng Du, Cui Tao, Hua Xu, and Yaoyun Zhang. 2020. Exploring temporal suicidal behavior patterns on social media: Insight from Twitter analytics. *Health informatics journal* 26, 2 (2020), 738–752.

[57] David D Luxton, Jennifer D June, and Jonathan M Fairall. 2012. Social media and suicide: a public health perspective. *American journal of public health* 102, S2 (2012), S195–S200.

[58] Yihua Ma and Youliang Cao. 2020. Dual Attention based Suicide Risk Detection on Social Media. In *2020 IEEE International Conference on Artificial Intelligence and Computer Applications (ICAICA)*. IEEE, 637–640.

[59] Naoki Masuda, Issei Kurahashi, and Hiroko Onari. 2013. Suicide ideation of individuals in online social networks. *PloS one* 8, 4 (2013), e62262.

[60] Matthew Matero, Akash Idnani, Youngseo Son, Salvatore Giorgi, Huy Vu, Mohammad Zamani, Parth Limbachiya, Sharath Chandra Guntuku, and H Andrew Schwartz. 2019. Suicide risk assessment with multi-level dual-context language and BERT. In *Proceedings of the Sixth Workshop on Computational Linguistics and Clinical Psychology*. 39–44.

[61] Diego Maupomé and Marie-Jean Meurs. 2018. Using Topic Extraction on Social Media Content for the Early Detection of Depression. *CLEF (Working Notes)* 2125 (2018).

[62] Catherine M McHugh, Amy Corderoy, Christopher James Ryan, Ian B Hickie, and Matthew Michael Large. 2019. Association between suicidal ideation and suicide: meta-analyses of odds ratios, sensitivity, specificity and positive predictive value. *BJPsych open* 5, 2 (2019).

[63] Leland McInnes, John Healy, and James Melville. 2018. Umap: Uniform manifold approximation and projection for dimension reduction. *arXiv preprint arXiv:1802.03426* (2018).

[64] Tomas Mikolov, Kai Chen, Greg Corrado, and Jeffrey Dean. 2013. Efficient estimation of word representations in vector space. *arXiv preprint arXiv:1301.3781* (2013).

[65] David N Milne, Glen Pink, Ben Hachey, and Rafael A Calvo. 2016. Clpsych 2016 shared task: Triaging content in online peer-support forums. In *Proceedings of the third workshop on computational linguistics and clinical psychology*. 118–127.

[66] Shervin Minaee, Nal Kalchbrenner, Erik Cambria, Narjes Nikzad, Meysam Chenaghlu, and Jianfeng Gao. 2021. Deep Learning–based Text Classification: A Comprehensive Review. *ACM Computing Surveys (CSUR)* 54, 3 (2021), 1–40.

[67] Rohan Mishra, Pradyumn Prakhar Sinha, Ramit Sawhney, Debanjan Mahata, Puneet Mathur, and Rajiv Ratn Shah. 2019. SNAP-BATNET: Cascading author profiling and social network graphs for suicide ideation detection on social media. In *Proceedings of the 2019 Conference of the North American Chapter of the Association for Computational Linguistics: Student Research Workshop*. 147–156.

[68] Margaret Mitchell, Kristy Hollingshead, and Glen Coppersmith. 2015. Quantifying the language of schizophrenia in social media. In *Proceedings of the 2nd workshop on Computational linguistics and clinical psychology: From linguistic signal to clinical reality*. 11–20.

[69] Seungwhan Moon, Leonardo Neves, and Vitor Carvalho. 2018. Multimodal named entity disambiguation for noisy social media posts. In *Proceedings of the 56th Annual Meeting of the Association for Computational Linguistics (Volume 1: Long Papers)*. 2000–2008.

[70] Michelle Morales, Stefan Scherer, and Rivka Levitan. 2017. A cross-modal review of indicators for depression detection systems. In *Proceedings of the fourth workshop on computational linguistics and clinical psychology—From linguistic signal to clinical reality*. 1–12.

[71] Bilel Moulahi, Jérôme Azé, and Sandra Bringay. 2017. DARE to care: a context-aware framework to track suicidal ideation on social media. In *International Conference on Web Information Systems Engineering*. Springer, 346–353.

[72] Hyeonseob Nam, Jung-Woo Ha, and Jeonghee Kim. 2017. Dual attention networks for multimodal reasoning and matching. In *Proceedings of the IEEE conference on computer vision and pattern recognition*. 299–307.

[73] Susan Nolen-Hoeksema. 1991. Responses to depression and their effects on the duration of depressive episodes. *Journal of abnormal psychology* 100, 4 (1991), 569.

[74] Bridianne O'dea, Mark E Larsen, Philip J Batterham, Alison L Calear, and Helen Christensen. 2017. A linguistic analysis of suicide-related Twitter posts. *Crisis: The Journal of Crisis Intervention and Suicide Prevention* 38, 5 (2017), 319.

[75] Yaakov Ophir, Refael Tikochinski, Christa SC Asterhan, Itay Sisso, and Roi Reichart. 2020. Deep neural networks detect suicide risk from textual facebook posts. *Scientific reports* 10, 1 (2020), 1–10.

[76] Ahmed Husseini Orabi, Prasadith Buddhitha, Mahmoud Husseini Orabi, and Diana Inkpen. 2018. Deep learning for depression detection of twitter users. In *Proceedings of the Fifth Workshop on Computational Linguistics and Clinical Psychology: From Keyboard to Clinic*. 88–97.

[77] Minsu Park, Chiyoung Cha, and Meeyoung Cha. 2012. Depressive moods of users portrayed in Twitter. (2012).

[78] Minsu Park, David McDonald, and Meeyoung Cha. 2013. Perception differences between the depressed and non-depressed users in twitter. In *Proceedings of the International AAAI Conference on Web and Social Media*, Vol. 7.

[79] James W Pennebaker, Ryan L Boyd, Kayla Jordan, and Kate Blackburn. 2015. *The development and psychometric properties of LIWC2015*. Technical Report.

[80] Inna Pirina and Çağrı Çöltekin. 2018. Identifying depression on reddit: The effect of training data. In *Proceedings of the 2018 EMNLP Workshop SMM4H: The 3rd Social Media Mining for Health Applications Workshop & Shared Task*. 9–12.

[81] Robert Plutchik. 1980. A general psychoevolutionary theory of emotion. In *Theories of emotion*. Elsevier, 3–33.

[82] Daniel Preoţiuc-Pietro, Johannes Eichstaedt, Gregory Park, Maarten Sap, Laura Smith, Victoria Tobolsky, H Andrew Schwartz, and Lyle Ungar. 2015. The role of personality, age, and gender in tweeting about mental illness. In *Proceedings of the 2nd workshop on computational linguistics and clinical psychology: From linguistic signal to clinical reality*. 21–30.

[83] Daniel Preotiuc-Pietro, Maarten Sap, H Andrew Schwartz, and Lyle H Ungar. 2015. Mental Illness Detection at the World Well-Being Project for the CLPsych 2015 Shared Task.. In *CLPsych@ HLT-NAACL*. 40–45.

[84] Tharindu Ranasinghe and Marcos Zampieri. 2021. Multilingual Offensive Language Identification for Low-resource Languages. *arXiv preprint arXiv:2105.05996* (2021).

[85] Andrew G Reece and Christopher M Danforth. 2017. Instagram photos reveal predictive markers of depression. *EPJ Data Science* 6 (2017), 1–12.

[86] Nils Reimers and Iryna Gurevych. 2019. Sentence-bert: Sentence embeddings using siamese bert-networks. *arXiv preprint arXiv:1908.10084* (2019).

[87] Philip Resnik, William Armstrong, Leonardo Claudino, and Thang Nguyen. 2015. The University of Maryland CLPsych 2015 shared task system. In *Proceedings of the 2nd workshop on computational linguistics and clinical psychology: from linguistic signal to clinical reality*. 54–60.

[88] Philip Resnik, William Armstrong, Leonardo Claudino, Thang Nguyen, Viet-An Nguyen, and Jordan Boyd-Graber. 2015. Beyond LDA: exploring supervised topic modeling for depression-related language in Twitter. In *Proceedings of the 2nd Workshop on Computational Linguistics and Clinical Psychology: From Linguistic Signal to Clinical Reality*. 99–107.

[89] Antoine Rolet, Marco Cuturi, and Gabriel Peyré. 2016. Fast dictionary learning with a smoothed Wasserstein loss. In *Artificial Intelligence and Statistics*. PMLR, 630–638.

[90] Arunima Roy, Katerina Nikolitch, Rachel McGinn, Safiya Jinah, William Klement, and Zachary A Kaminsky. 2020. A machine learning approach predicts future risk to suicidal ideation from social media data. *NPJ digital medicine* 3, 1 (2020), 1–12.

[91] Olga Russakovsky, Jia Deng, Hao Su, Jonathan Krause, Sanjeev Satheesh, Sean Ma, Zhiheng Huang, Andrej Karpathy, Aditya Khosla, Michael Bernstein, et al. 2015. Imagenet large scale visual recognition challenge. *International journal of computer vision* 115, 3 (2015), 211–252.

[92] Elvis Saravia, Chun-Hao Chang, Renaud Jollet De Lorenzo, and Yi-Shin Chen. 2016. MIDAS: Mental illness detection and analysis via social media. In *2016 IEEE/ACM International Conference on Advances in Social Networks Analysis and Mining (ASONAM)*. IEEE, 1418–1421.

[93] Ramit Sawhney, Harshit Joshi, Lucie Flek, and Rajiv Shah. 2021. PHASE: Learning Emotional Phase-aware Representations for Suicide Ideation Detection on Social Media. In *Proceedings of the 16th Conference of the European Chapter of the Association for Computational Linguistics: Main Volume*. 2415–2428.

[94] Ramit Sawhney, Harshit Joshi, Saumya Gandhi, and Rajiv Ratn Shah. 2021. Towards Ordinal Suicide Ideation Detection on Social Media. In *Proceedings of the 14th ACM International Conference on Web Search and Data Mining*. 22–30.

[95] Ramit Sawhney, Harshit Joshi, Rajiv Shah, and Lucie Flek. 2021. Suicide Ideation Detection via Social and Temporal User Representations using Hyperbolic Learning. In *Proceedings of the 2021 Conference of the North American Chapter of the Association for Computational Linguistics: Human Language Technologies*. 2176–2190.

[96] Ramit Sawhney, Prachi Manchanda, Puneet Mathur, Rajiv Shah, and Raj Singh. 2018. Exploring and learning suicidal ideation connotations on social media with deep learning. In *Proceedings of the 9th Workshop on Computational Approaches to Subjectivity, Sentiment and Social Media Analysis*. 167–175.

[97] Ramit Sawhney, Prachi Manchanda, Raj Singh, and Swati Aggarwal. 2018. A computational approach to feature extraction for identification of suicidal ideation in tweets. In *Proceedings of ACL 2018, Student Research Workshop*. 91–98.

[98] Harold Schlosberg. 1954. Three dimensions of emotion. *Psychological review* 61, 2 (1954), 81.

[99] H Andrew Schwartz, Salvatore Giorgi, Maarten Sap, Patrick Crutchley, Lyle Ungar, and Johannes Eichstaedt. 2017. Dlatk: Differential language analysis toolkit. In *Proceedings of the 2017 conference on empirical methods in natural language processing: System demonstrations*. 55–60.

[100] Alfonso Semeraro, Salvatore Vilella, and Giancarlo Ruffo. 2021. PyPlutchik: visualising and comparing emotion-annotated corpora. *arXiv preprint arXiv:2105.04295* (2021).

[101] Faisal Muhammad Shah, Farsheed Haque, Ragib Un Nur, Shaeekh Al Jahan, and Zarar Mamud. 2020. A Hybridized Feature Extraction Approach To Suicidal Ideation Detection From Social Media Post. In *2020 IEEE Region 10 Symposium (TENSYMP)*. IEEE, 985–988.

[102] Guangyao Shen, Jia Jia, Liqiang Nie, Fuli Feng, Cunjun Zhang, Tianrui Hu, Tat-Seng Chua, and Wenwu Zhu. 2017. Depression Detection via Harvesting Social Media: A Multimodal Dictionary Learning Solution.. In *IJCAI*. 3838–3844.

[103] Tiancheng Shen, Jia Jia, Guangyao Shen, Fuli Feng, Xiangnan He, Huanbo Luan, Jie Tang, Thanassis Tiropanis, Tat Seng Chua, and Wendy Hall. 2018. Cross-domain depression detection via harvesting social media. International Joint Conferences on Artificial Intelligence.

[104] Han-Chin Shing, Philip Resnik, and Douglas W Oard. 2020. A prioritization model for suicidality risk assessment. In *Proceedings of the 58th Annual Meeting of the Association for Computational Linguistics*. 8124–8137.

[105] Karen Simonyan and Andrew Zisserman. 2014. Very deep convolutional networks for large-scale image recognition. *arXiv preprint arXiv:1409.1556* (2014).

[106] Pradyumna Prakhar Sinha, Rohan Mishra, Ramit Sawhney, Debanjan Mahata, Rajiv Ratn Shah, and Huan Liu. 2019. # suicidal-A multipronged approach to identify and explore suicidal ideation in twitter. In *Proceedings of the 28th ACM International Conference on Information and Knowledge Management*. 941–950.

[107] Hoyun Song, Jinseon You, Jin-Woo Chung, and Jong C Park. 2018. Feature Attention Network: Interpretable Depression Detection from Social Media.. In *PACLIC*.

[108] Xuemeng Song, Liqiang Nie, Luming Zhang, Mohammad Akbari, and Tat-Seng Chua. 2015. Multiple social network learning and its application in volunteerism tendency prediction. In *Proceedings of the 38th International ACM SIGIR Conference on Research and Development in Information Retrieval*. 213–222.

[109] Maxim Stankevich, Vadim Isakov, Dmitry Devyatkin, and Ivan Smirnov. 2018. Feature Engineering for Depression Detection in Social Media.. In *ICPRAM*. 426–431.

[110] Deborah M Stone. 2021. Changes in Suicide Rates—United States, 2018–2019. *MMWR. Morbidity and Mortality Weekly Report* 70 (2021).

[111] Michael M Tadesse, Hongfei Lin, Bo Xu, and Liang Yang. 2019. Detection of depression-related posts in reddit social media forum. *IEEE Access* 7 (2019), 44883–44893.

[112] Michael Mesfin Tadesse, Hongfei Lin, Bo Xu, and Liang Yang. 2020. Detection of suicide ideation in social media forums using deep learning. *Algorithms* 13, 1 (2020), 7.

[113] Anja Thieme, Danielle Belgrave, and Gavin Doherty. 2020. Machine learning in mental health: A systematic review of the HCI literature to support the development of effective and implementable ML systems. *ACM Transactions on Computer-Human Interaction (TOCHI)* 27, 5 (2020), 1–53.

[114] Sho Tsugawa, Yusuke Kikuchi, Fumio Kishino, Kosuke Nakajima, Yuichi Itoh, and Hiroyuki Ohsaki. 2015. Recognizing depression from twitter activity. In *Proceedings of the 33rd annual ACM conference on human factors in computing systems*. 3187–3196.

[115] Elsbeth Turcan and Kathleen McKeown. 2019. Dreaddit: A Reddit dataset for stress analysis in social media. *arXiv preprint arXiv:1911.00133* (2019).

[116] Elsbeth Turcan, Smaranda Muresan, and Kathleen McKeown. 2021. Emotion-Infused Models for Explainable Psychological Stress Detection. In *Proceedings of the 2021 Conference of the North American Chapter of the Association for Computational Linguistics: Human Language Technologies*.

2895–2909.

[117] Nikhita Vedula and Srinivasan Parthasarathy. 2017. Emotional and linguistic cues of depression from social media. In *Proceedings of the 2017 International Conference on Digital Health*. 127–136.

[118] James Vincent. 2017. Facebook is using AI to spot users with suicidal thoughts and send them help. *The Verge* (2017).

[119] M Johnson Vioules, Bilel Moulahi, Jérôme Azé, and Sandra Bringay. 2018. Detection of suicide-related posts in Twitter data streams. *IBM Journal of Research and Development* 62, 1 (2018), 7–1.

[120] Ning Wang, Fan Luo, Yuvraj Shivtare, Varsha D Badal, KP Subbalakshmi, Rajarathnam Chandramouli, and Ellen Lee. 2021. Learning Models for Suicide Prediction from Social Media Posts. *arXiv preprint arXiv:2105.03315* (2021).

[121] Wei Wang, Yan Huang, Yizhou Wang, and Liang Wang. 2014. Generalized autoencoder: A neural network framework for dimensionality reduction. In *Proceedings of the IEEE conference on computer vision and pattern recognition workshops*. 490–497.

[122] Yilin Wang, Jiliang Tang, Jundong Li, Baoxin Li, Yali Wan, Clayton Mellina, Neil O'Hare, and Yi Chang. 2017. Understanding and discovering deliberate self-harm content in social media. In *Proceedings of the 26th International Conference on World Wide Web*. 93–102.

[123] Cynthia Whissell. 2009. Using the revised dictionary of affect in language to quantify the emotional undertones of samples of natural language. *Psychological reports* 105, 2 (2009), 509–521.

[124] Owen Whooley. 2014. Diagnostic and statistical manual of mental disorders (DSM). *The Wiley Blackwell Encyclopedia of Health, Illness, Behavior, and Society* (2014), 381–384.

[125] Akkapon Wongkoblap, Miguel A Vadillo, and Vasa Curcin. 2017. Researching mental health disorders in the era of social media: systematic review. *Journal of medical Internet research* 19, 6 (2017), e228.

[126] Jun-Ming Xu, Kwang-Sung Jun, Xiaojin Zhu, and Amy Bellmore. 2012. Learning from bullying traces in social media. In *Proceedings of the 2012 conference of the North American chapter of the association for computational linguistics: Human language technologies*. 656–666.

[127] Zhentao Xu, Verónica Pérez-Rosas, and Rada Mihalcea. 2020. Inferring Social Media Users' Mental Health Status from Multimodal Information. In *Proceedings of The 12th Language Resources and Evaluation Conference*. 6292–6299.

[128] Zichao Yang, Diyi Yang, Chris Dyer, Xiaodong He, Alex Smola, and Eduard Hovy. 2016. Hierarchical attention networks for document classification. In *Proceedings of the 2016 conference of the North American chapter of the association for computational linguistics: human language technologies*. 1480–1489.

[129] Andrew Yates, Arman Cohan, and Nazli Goharian. 2017. Depression and self-harm risk assessment in online forums. *arXiv preprint arXiv:1709.01848* (2017).

[130] Amir Hossein Yazdavar, Mohammad Saeid Mahdavinejad, Goonmeet Bajaj, William Romine, Amit Sheth, Amir Hassan Monadjemi, Krishnaprasad Thirunarayan, John M Meddar, Annie Myers, Jyotishman Pathak, et al. 2020. Multimodal mental health analysis in social media. *Plos one* 15, 4 (2020), e0226248.

[131] Sicheng Zhao, Shangfei Wang, Mohammad Soleymani, Dhiraj Joshi, and Qiang Ji. 2019. Affective computing for large-scale heterogeneous multimedia data: A survey. *ACM Transactions on Multimedia Computing, Communications, and Applications (TOMM)* 15, 3s (2019), 1–32.

[132] Yiheng Zhou, Jingyao Zhan, and Jiebo Luo. 2017. Predicting multiple risky behaviors via multimedia content. In *International Conference on Social Informatics*. Springer, 65–73.

[133] Hamad Zogan, Imran Razzak, Shoaib Jameel, and Guandong Xu. 2021. DepressionNet: A Novel Summarization Boosted Deep Framework for Depression Detection on Social Media. *arXiv preprint arXiv:2105.10878* (2021).

[134] Hamad Zogan, Imran Razzak, Xianzhi Wang, Shoaib Jameel, and Guandong Xu. 2020. Explainable Depression Detection with Multi-Modalities Using a Hybrid Deep Learning Model on Social Media. *arXiv preprint arXiv:2007.02847* (2020).

[135] Hamad Zogan, Xianzhi Wang, Shoaib Jameel, and Guandong Xu. 2020. Depression detection with multi-modalities using a hybrid deep learning model on social media. *arXiv preprint arXiv:2007.02847* (2020).